\documentclass[a4paper,12pt,twoside]{article}
\usepackage{amsfonts}
\usepackage{amsmath}
\usepackage{amssymb}
\usepackage{amsthm}

\newcommand{\bm}[1]{\mbox{\boldmath $#1$}}

\usepackage{graphicx,amssymb}
\usepackage{color}
\usepackage{bm}

\newcounter{mnotecount}

\newcommand{\mnotex}[1]
{\protect{\stepcounter{mnotecount}}$^{\mbox{\footnotesize $\bullet$\themnotecount}}$ 
\marginpar{
\raggedright\tiny\em
$\!\!\!\!\!\!\,\bullet$\themnotecount: #1} }

\def\CQG{Class. Quantum Grav.}
\def\PRD{Phys. Rev. D}
\def\GRG{Gen. Rel. Grav.}
\def\PHYSREV{Phys. Rev.}
\def\RMP{Rev. Mod. Phys.}
\def\MNRAS{Mon. Not. Roy. Astr. Soc.}
\def\JMP{J. Math. Phys.}
\def\CMP{Comm. Math. Phys.}
\def\PRL{ Phys. Rev. Lett.}
\def\APJ{Astrophys. J.}
\def\AANDA{Astron. Astrophys.}
\def\AASFSA{Ann. Acad. Soc. Sci. Fennicae Ser. A}
\def\ZP{Z. Phys.}
\def\PLB{Phys. Lett. B}
\def\ATMP{Adv. Theor. Math. Phys.}
\def\COMRENP{C. R. Phys.}
\def\PPNP{Prog. Part. Nucl. Phys.}

\def\p{\partial}
\def\sup{\Omega}

\def\eqq{\stackrel{\sup}{=}}
\def\mmm{\mathcal{V}}
\def\mm2{\mathcal{W}}
\def\mmextended{\widehat{\mathcal{W}}} 
\def\embed{\Psi}

\def\FL{\mathrm{\scriptscriptstyle RW}}
\def\HOM{\mathrm{\scriptscriptstyle HOM}}
\def\sx{\mathrm{\scriptscriptstyle SX}}
\def\stc{\mathrm{\scriptscriptstyle ST}}

\def\spsx{(\mmm^{\sx}, g^{\sx})}

\def\flr{(\mm2^{\FL}, g^{\FL})}
\def\sxr{(\mm2^{\sx}, g^{\sx})}
\def\str{(\mm2^{\stc}, g^{\stc})}

\def\rhofl{\rho^{\FL}}
\def\prefl{p^{\FL}}

\def\M{{\cal M}}
\def\Sch{\mathrm{\scriptscriptstyle Sch}}
\def\K{\mathrm{\scriptscriptstyle Kot}}

\def\tphi{\Phi}

\def\ax{\eta}
\def\axfl{\eta_{\FL}}

\def\stk{\xi}

\def\lie{\mathcal{L}}

\definecolor{kernel}{rgb}{0.4,0.4,0.4}

\def\sig{\Sigma}
\def\dersigo{{\Sigma'_c}}
\def\sigo{\Sigma_c}
\def\dersig{\Sigma'}
\def\curv{\epsilon}              

\def\cy{Y}
\def\ppi{\Pi}
\def\cchi{\chi}
\def\ch{{\cal H}}

\def\yone{{{\cal Y}_1}}
\def\ytwo{{{\cal Y}_2}}
\def\wone{{{\cal W}_1}}
\def\wzero{W}
\def\wtwo{{{\cal W}_2}}
\def\qone{{{\cal Q}_1}}
\def\qtwo{{{\cal Q}_2}}
\def\uone{{{\cal U}_1}}
\def\utwo{{{\cal U}_2}}

\def\asig{a}
\def\asigdot{\dot a}

\def\derxiutwo{{{\cal U}_2}'}

\def\cg{{\cal G}}

\def\rsig{r_0}
\def\rsigdot{\dot{r}_0}
\def\rsigdotdot{\ddot{r}_0}
\def\tsig{T_0}
\def\tsigdot{\dot{T}_0}
\def\tsigdotdot{\ddot{T}_0}

\def\cgsup{{\cal G}}

\def\der2cpsupr{ \frac{\partial^2 {\cal P}}{\partial r^2} }
\def\dercgsupr{ \frac{\partial {\cal G}}{\partial r} }

\def\K{\mathcal{K}}
\def\gback{g^{(0)}}

\def\gfpert{g^{(1)}}

\def\supo{\sup_0}

\def\eqq{\stackrel{\sup_0}{=}}

\def\wwtwo{{{\cal W}_2} 
-a [{\dot{\cal U}_2} + {\dot {\cal Y}_2}
]
}

\def\fff{q}
\def\sff{K}
\def\nback{n^{(0)}}

\def\fffback{\fff^{(0)}}
\def\ffffpert{\fff^{(1)}}
\def\sffback{\sff^{(0)}}
\def\sfffpert{\sff^{(1)}}

\def\defi{:=}

\def\hodge{\stackrel{\mathbb{S}^2}{\to}}

\def\tra{{tr}}

\def\X{{\cal X}}
\def\Z{{\cal Z}}

\def\ZStwo{{\Z}^{\mathbb{S}^2}}

\def\Kas{\mathcal{K}}

\begin{document}

\title{
Review on exact and perturbative deformations of the Einstein-Straus
model: uniqueness and rigidity results
%
} 

\author{Marc Mars\footnotemark[1], 
Filipe C. Mena\footnotemark[2] and
Ra\"ul Vera\footnotemark[3] \\
{\small \footnotemark[1] Instituto de F\'{\i}sica Fundamental y Matem\'aticas,
Universidad de Salamanca,} \\
{\small Plaza de la Merced s/n, 37008 Salamanca, Spain} \\
{\small \footnotemark[2] Centro de Matem\'atica, Universidade do Minho, 4710-057 Braga,
Portugal}
\\
{\small \footnotemark[3] Dept. of Theoretical Physics and History of Science,}
\\ {\small University of the Basque Country UPV/EHU,
644 PK, Bilbao 48080, Basque Country, Spain }}

\date{}
\maketitle

\begin{abstract}
The Einstein-Straus model consists of 
a Schwarzschild spherical vacuole in a Friedman-Lema\^{\i}tre-Robertson-Walker (FLRW) dust
spacetime (with or without $\Lambda$).
It constitutes the 
most widely accepted model
to answer the question of the influence of large scale (cosmological)
dynamics on local systems. The conclusion drawn by the model is that
there is no influence from the cosmic background, since the spherical
vacuole is static. Spherical generalizations to other interior matter models
are commonly used in the construction of lumpy inhomogeneous
cosmological models. On the other hand, the model has proven to be reluctant
to admit non-spherical generalizations. In this review, we summarize
the known uniqueness results for this model. These seem to indicate that 
the only reasonable and realistic
non-spherical deformations of the Einstein-Straus
model require perturbing the FLRW background. We review results about
linear perturbations of the Einstein-Straus model, where the perturbations
in the vacuole are assumed to be stationary and axially symmetric so as to
describe regions (voids in particular)
in which the matter has reached an equilibrium regime.

\end{abstract}

\section{Introduction}

During a meal in the 19th Jena Meeting on Relativity, in September
1996, Bill Bonnor provocatively asked Jos\'e Senovilla if the table
could be expanding with the Universe. Not surprisingly, Bonnor later
took the question seriously and wrote a paper about how the hydrogen
atom is affected by the cosmic expansion \cite{bonnor99}, which is
well worth reading.

About five decades before, Einstein and Straus asked a similar question, on a
bigger scale, which led them to investigate {\em the
influence of the expansion of space on the gravitational fields
surrounding individual stars} \cite{Einstein45}. They took the
Schwarzschild solution representing the vacuole surrounding a star
located in the centre and the Friedman-Lemaitre-Robertson-Walker
(FLRW) solution as the cosmological model. At the core of their model
was the matching of the two solutions across a spherical surface with
constant cosmological radius. Since the expansion kept the vacuole symmetry and
time independence, 
their conclusion was that it does not affect the
gravitational fields surrounding stars and, in particular, it does not
affect the solar system dynamics. 
A previous attempt to address the issue
of whether the planetary orbits expand with the Universe 
was made by McVittie \cite{McVittie} who found a smooth 
model describing a spherically symmetric mass embedded in a flat FLRW.

Since then the research about this problem was
scarce, although some alternatives to the McVittie model
were suggested, e.g. in \cite{Gautreau84}, and difficulties of the global
meaning of the model were also pointed out \cite{Nolan93,
Nolan98, Nolan99a, Nolan99b} (see also \cite{Carrera}).
Concerning the Einstein-Straus model itself,
it was revisited in \cite{Balbinot88, Bona87}
and stability issues were raised
in \cite{Sato84},  \cite{LakePim85} cf. also the discussion in 
\cite{Krasinski}. However, the Einstein-Straus model has never
stopped being 
considered as the correct answer to the lack of influence of the cosmological
expansion on local systems. Moreover, since the vacuole can be inserted anywhere
due to the homogeneity of FLRW, the Einstein-Straus model led to the original Swiss cheese
model of a lumpy inhomogeneous universe (see e.g. \cite{RelCosm}).

Bonnor's question, that 1996
afternoon, raised a totally new
issue for the Einstein-Straus models, namely whether spherical
symmetry was a crucial ingredient of the model and, therefore, for
the existence of time-invariant bounded systems embedded in 
a FLRW universe.  
Indeed, the question triggered research
by Senovilla and Vera \cite{Seno_Vera_97},
that led to the result about the impossibility of the Einstein-Straus
model in cylindrical symmetry. In turn, this important result was the
origin of over fifteen years of research about the rigidity,
in the sense of uniqueness, of the model. The aim of this paper is to
review these results on rigidity both for exact models and from a
perturbative perspective.

Crucial to this endeavour was the development of a general
mathematical theory of spacetime matching \cite{Mars-Seno_99} and of
its perturbative version \cite{BattyeCarter01, Mukohyama00,
Mars05}. This allowed to achieve quite
general results about the possibility of generalizing both the shape
of the cavity and the cosmological setting of the original
Einstein-Straus model. As an example, described below in some detail, 
uniqueness of the static $\Lambda$-vacuum spherical region
embedded in a non-static FLRW cosmological model has been proved
\cite{MarsES2}.  

The scope of the Einstein-Straus model has been taken well
beyond  both the physical scale
originally considered 
and the physical problems for which the model was conceived.
In fact, the model has been used not only at the solar system scale,
but also on galaxy \cite{Ishak08} and galaxy clusters' scales
\cite{Ishak08,Plaga05}.
On the other hand, the vacuole of the (original) Einstein-Straus model
has been replaced by other spherically symmetric geometries, generally
Lema\^itre-Tolman-Bondi (LTB), also spherically shaped regions of Szekeres,
in order to construct {\em ``generalized'' Einstein-Straus}
models for describing extra-galactic scale and cosmic voids 
(we refer to the reviews in \cite{Krasinski,RelCosm}).
Lumpy inhomogeneous cosmological models
based on the generalized Einstein-Straus Swiss cheese models are being used
in the search of possible explanations
to the accelerated expansion of the Universe 
by the study of lensing effects at cosmic scales produced by the voids (see e.g. \cite{RelCosm}).

So far, all these generalized Einstein-Straus (and the corresponding
Swiss cheese) models have assumed spherically shaped inhomogeneities (voids).
One of 
themes of the research we will review here is how far can one push
the Einstein-Straus model towards non-spherical generalizations. The fundamental
ingredient we want to keep is that the bound system remains stationary, so
as to keep the absence of influence of the cosmic dynamics on 
astrophysical scales. We will use the term  {\em Einstein-Straus problem}
to the problem of finding the most general stationary regions (vacuum or not)
one can embed in a realistic cosmological model in a broad sense.

The Einstein-Straus model has also been taken beyond the 
exact solutions' settings to
include metric perturbations. Perturbation theory in General
Relativity (GR) is a natural framework to study small departures from
symmetric configurations and thus to perform stability analysis. For
instance, it allows to include simultaneously density, rotational and
gravitational wave perturbation modes into an, otherwise, spatially
homogeneous and isotropic cosmological model. Most interestingly, it
allows {\em a priori} to perturb independently the interior and
exterior spacetimes as well as the matching boundary. Furthermore,
although the three perturbation modes are decoupled on a FLRW
background, they may couple at a matching boundary. In this context,
an important question is how general can the perturbations be in each
model.

Inherent to  perturbation theory is the issue of gauge
freedom. In perturbed spacetime matchings, this can be
complicated by the fact that three independent perturbation gauges
may be in use. For the sake of completeness, we include
a short review of linearized matching, where these issues
are discussed, see also \cite{Mukohyama00} for further
details, including the definition of the 
so-called {\em doubly gauge invariant variables}.

Perturbations in the FLRW background are customarily
split in scalar, vector and tensor modes, and the later 
are generically viewed as cosmological gravitational waves.
Given that the gravitational wave detectors are already
active, and gravitational waves are expected to be detected within the
next five years (see e.g. \cite{Papa13, Riles13} for recent accounts
and the review \cite{Living_Review_GW}), it would be interesting to investigate
their inclusion in the models. One now certainly has the necessary
mathematical machinery to do so, and preliminary results 
indicate the possibility of having, for instance, a stationary axially
symmetric vacuole embedded in an expanding cosmological model
containing tensor modes \cite{ESpert2}. Even more,
the linearized matching links the rotational and tensor modes
degrees of freedom in the perturbation variables 
\cite{ESpert2}.

This is therefore an interesting timing to revise the state-of-the-art
and point out potentially interesting directions of research about the
Einstein-Straus problem, in the sense pointed out above.

The plan of the review is the following. In Section \ref{sec:ESmodel}, 
the Einstein-Straus
model is briefly presented. Similar summaries with different degrees
of detail can be found in many places in the literature,
see e.g. \cite{RelCosm}. We include it here for the sake of completeness
and in order to fix our notation.  Section 
\ref{sec_rigidity} is devoted to describing in some detail the uniqueness
results concerning both general static regions and stationary and axisymmetric
regions (irrespective of any symmetry consideration
and/or matter content) embeddable in a FLRW expanding cosmology.
Section \ref{sec:STATIC} is devoted to the static case. The main
conclusion here is that the only static vacuum region that can be embedded
to an expanding FLRW is a spherically shaped region of Schwarzschild (i.e.
the Einstein-Straus model). Similar uniqueness results hold for other matter
models, such as vacuum with cosmological constant.
Section \ref{sec:STAX} deals, in turn, with the uniqueness results
for stationary and axially symmetric regions in FLRW expanding universes.
The main result is that the stationary region must, in fact, be static, so that the
previous conclusions on static regions apply. The uniqueness result
thus states that the only way of having a stationary and axially symmetric
or static region in an expanding FLRW is the Einstein-Straus model.
Following the uniqueness results, the robustness of the
Einstein-Straus model is further analyzed by considering  
alternative exact cosmological models. In Section 
\ref{sec:robustness} the replacement of the FLRW region
by more general anisotropic cosmologies, i.e. the Bianchi models,
is studied for static locally cylindrically symmetric interiors,
leading to severe restrictions and no-go results for reasonable
evolving cosmologies.
The final part of the paper is devoted to
the generalization of the Einstein-Straus model
from a perturbative perspective. After a brief overview of perturbative
matching theory in Section \ref{sec:overview}, and the use of the Hodge
decomposition on the sphere instead of the usual spherical harmonic
decomposition in Section \ref{sec:hodge}, the linearized matching
between stationary and axisymmetric perturbations of Schwarzschild
and general perturbed FLRW is reviewed in Section \ref{sec:linearizedmatching}.
We finish with some conclusions in Section \ref{sec:conclusions},
pointing out some ongoing research,
prospects for future work on the perturbed Einstein-Straus model
and its possible implications on the relationship between
astrophysical bounded systems and cosmological dynamics
in the form of cosmic gravitational waves.

\section{The Einstein-Straus model}
\label{sec:ESmodel}

This model consists of a spherically symmetric (both in shape
and intrinsic geometry) Ricci-flat region embedded in a FLRW universe
without cosmological constant.
Recall that two spacetimes can be matched
across their boundary if and only if the first fundamental
forms $q$ and second
fundamental forms $K$ agree on the matching hypersurface. A consequence
of this are the well-known Israel conditions, which restrict
the jump of the energy-momentum tensor
across the boundary. In the present context, they
imply that the cosmological fluid must be dust
and the  vacuole must be comoving with the cosmological fluid. 
Writing the FLRW metric in cosmic time coordinates
\begin{eqnarray}
g^{\FL} = -dt^2 + a^2 (t) g_{\M}, \qquad \mbox{with }\qquad
g_{\M}= dR^2 +\Sigma^2(R,\epsilon) d \Omega^2, 
\label{FLmetric}
\end{eqnarray}
where $\epsilon = \{ -1,0,+1\} $, ${\Sigma'}{}^2 = 1- \epsilon \Sigma^2$
with prime denoting derivative with respect to $R$, in units $G=c=1$
the Friedman equation reads
\begin{eqnarray*}
\dot{a}^2  + \epsilon  =
\frac{8 \pi \rho_0}{3a},
\end{eqnarray*}
where the dot denotes derivative with respect to $t$ and $\rho_0$ is
a constant such that the cosmological
energy-density $\rho$ satisfies $\rho = \rho_0/a^3$.
The boundary of the vacuole can be 
parametrized by $\{ t = t, R= R_0 \}$ (we ignore
the angular variables as they behave trivially, and use $t$ both as spacetime
coordinate and intrinsic coordinate on the hypersurface, the 
precise meaning will be clear from the context).
For the matching one needs
the induced metric $q^{\FL}$ and the second fundamental form $K^{\FL}$.
Using the outward unit normal $\bm{n}_{\FL} = a(t) dR$ these
objects read, with $\Sigma_c \defi \Sigma |_{R=R_0}$, and $\Sigma'_c 
\defi \Sigma'|_{R=R_0}$,
\begin{eqnarray*}
q^{\FL} = -dt^2 + a^2(t)  \Sigma_c^2 d \Omega^2, \quad \quad
K^{\FL} = a(t)  \Sigma_c \Sigma'_c d \Omega^2.
\end{eqnarray*}
From  Birkhoff's theorem, the geometry of the vacuole is Kruskal.
Assuming that the boundary is away from
the Schwarzschild horizon (this happens sufficiently away from the big
bang or big crunch) the interior metric can be written in
Schwarzschild coordinates
\begin{equation}
\label{SCHmetric}
g^{\Sch} = - \left ( 1 - \frac{2m}{r} \right )^2 dT^2 + 
\frac{dr^2}{1 - \frac{2m}{r}} + r^2 d\Omega^2.
\end{equation}
The boundary can be parametrized as $\{ T = \tsig (t), r = \rsig (t) \}$
and, given the time inversion symmetry of Schwarzschild, we can assume
$\tsigdot >0$ without loss of generality. The induced metric on the boundary is
\begin{eqnarray*}
q^{\Sch} = - N^2(t) dt^2 + \rsig^2(t) d \Omega^2, \quad \quad
N^2 = \left ( 1 - \frac{2m}{\rsig} \right ) \tsigdot^2
- \frac{\rsigdot^2}{1 - \frac{2m}{\rsig}}.
\end{eqnarray*}
Using the unit normal $\bm{n}_{\Sch} = \frac{1}{N} 
( \tsigdot dr - \rsigdot dT )$ (note that the global sign of $N$ is kept free
at this stage), the second fundamental form is
\begin{eqnarray*}
K^{\Sch} = \frac{1}{N} \left (  -\tsigdot \rsigdotdot 
+ \tsigdotdot  \rsigdot + \frac{3m \rsigdot^2 \tsigdot}{\rsig \left (\rsig 
- 2m \right )} - \frac{m}{\rsig^2} \left ( 1- \frac{2m}{\rsig} \right ) \tsigdot^3
\right ) dt^2 + \frac{\tsigdot \left (\rsig  -2 m \right )}{N} d\Omega^2.
\end{eqnarray*}
Equality of the $t$-component of the induced metric requires $N^2=1$. Then,
equality of the angular parts of the first and second fundamental forms imposes
$N=1$ and
\begin{eqnarray}
\rsig (t) = \Sigma_c a(t), \quad \quad
\tsigdot = \frac{a(t) \Sigma_c \Sigma'_c}{a (t) \Sigma_c - 2 m }.
\label{matchingODE}
\end{eqnarray}
A straightforward computation shows that 
the equality of the $t$-component of the 
induced metric
and second fundamental forms 
are  satisfied provided
the values of $\rho_0$
and $m$ are linked by $$m = \frac{4 \pi}{3} \rho_0 \Sigma_c^3,$$
which  has a clear interpretation in terms of (Misner-Sharp)-mass conservation.
This is the Einstein-Straus model \cite{Einstein45}.

A natural generalization consists in adding a cosmological
constant $\Lambda$ both to the FLRW and to the interior part (originally
considered in \cite{ES_with_Lambda_Gilbert} and fully solved in
\cite{Balbinot88}).
The  Israel matching conditions on the energy-momentum tensor now impose
the FLRW matter model to be dust with $\Lambda$, so that the Friedman equation is now
\begin{eqnarray*}
\dot{a}^2 + \epsilon  =
\frac{8 \pi \rho_0}{3a}  + \frac{\Lambda}{3} a^2.
\end{eqnarray*}
By Birkhoff's theorem, the interior metric is the Kottler solution (also
known as ``Schwarzschild- (anti) de Sitter''), which away from the horizons
is
\begin{eqnarray*}
g^{\K} = - \left ( 1 - \frac{2m}{r} - \frac{\Lambda r^2}{3} 
\right )^2 dT^2 + 
\frac{dr^2}{1 - \frac{2m}{r} - \frac{\Lambda r^2}{3} } + r^2 d\Omega^2.
\end{eqnarray*}
In this case, the matching conditions are
\begin{eqnarray*}
\rsig(t)  = \Sigma_c a(t) , \quad
\quad 
\tsigdot = \frac{a(t) \Sigma_c \Sigma'_c}{a (t) \Sigma_c - 2 m  
  - \frac{\Lambda}{3} \Sigma_c^3 a(t)^3},  \quad \quad 
m = \frac{4 \pi}{3} \rho_0 \Sigma_c^3.
\end{eqnarray*}
We emphasize that any matching of two spacetimes immediately leads to 
a complementary matching, at least locally, where the ``interior''
and ``exterior'' regions on each spacetime
reverse their roles (see \cite{FST_matching_spher_96} for details).
In the matching above, this leads to the Oppenheimer-Snyder collapse model \cite{OPSN}.

\section{Rigidity of the Einstein-Straus model}
\label{sec_rigidity}

The Einstein-Straus model is such that, for a given total mass
inside the vacuole and a given energy density in the cosmological
background, the radius of the static region is uniquely fixed.
This already poses difficulties for the model since it is often the case
that the size of the vacuole does not match the observed sizes 
of clustered matter in the universe, such as  stars or galaxies. This fact
indicates that the Einstein-Straus model may be lacking flexibility to
accommodate the various situations present in cosmology (cf. 
\cite{Schucking} and 
\cite{Bonnor96}).
In fact, the Einstein-Straus vacuole was found to be radially unstable in a
certain sense \cite{Krasinski}.
The other main
restriction of the model is its exact spherical symmetry. It is clear that
vacuoles in the universe are not exactly spherically symmetric so a natural
question is how robust is the model to non-spherical generalizations.

The first thing to consider is which fundamental ingredients of the
model should be kept. The main motivation of the Einstein-Straus model
was its ability to combine
cosmological expansion at large scales with no observable effects on the local
physics. Thus, the fundamental ingredient to keep is the absence of influence
of the cosmic expansion inside the region. The simplest and most natural
way to achieve this is imposing that the interior geometry
is stationary, because then no dynamical effects whatsoever from the 
surrounding evolving cosmology would affect the local physics. 
Among stationary interiors, the simplest case corresponds to static
situations, so it is natural to start with this case  (note that 
the Einstein-Straus model is itself static).

The question is then how rigid or flexible is the possibility
of having stationary/static regions embedded in an otherwise expanding  FLRW
universe. Ideally, one would like to make no further assumptions and find
the most general model with these properties. The matter model inside
may also be kept arbitrary, and see what are the possibilities allowed by 
the coexistence of a stationary/static region inside an expanding
FLRW universe. This coexistence has been sometimes called the Einstein-Straus \emph{problem}
in the literature
(see e.g. \cite{Balbinot88}). 
The first indication that the Einstein-Straus model might be very rigid
came from a seminal work by Senovilla and Vera \cite{Seno_Vera_97} who
considered the possibility of matching a static and cylindrically symmetric
region (with no restriction on the matter model) with 
a FLRW dynamical cosmology.
The matching hypersurface was
taken to be \emph{locally} a cylinder, in the sense of
being tangent in an open set
to the two (commuting) generators of two spatial local isometries.
No global consideration was needed.
The result was a no-go theorem: no such model exists.

With the impossibility of generalizing the Einstein-Straus model to
a cylindrical setting, it became of interest to study the problem
in as much generality as possible. The static case was treated in complete
generality in \cite{MarsES1}, \cite{MarsES2} and it is by now well-understood.
The more complicated stationary situation has been studied \cite{Nolan_Vera_05}
under the additional
assumptions of axisymmetry and a group action orthogonally transitive.
The motivation to study this simplified problem
lies in the fact that one expects equilibrium configurations to also exhibit
an axial symmetry. 
It is worth to mention that one step in the black hole uniqueness
theorems corresponds to showing that the domain of outer communications
must be axially symmetric 
\cite{Hawking-Ellis}.

We devote the following Section \ref{sec:STATIC} to describing the main results
in the static setting, and Section \ref{sec:STAX} to review the 
uniqueness results in the stationary and axially symmetric setting.

\subsection{Uniqueness results in the static case}
\label{sec:STATIC}

This case was first considered in \cite{MarsES1} under the 
additional assumption of axial symmetry, and one extra technical
assumption relating cosmic and static times on the matching hypersurfaces.
Both assumptions were dropped  in \cite{MarsES2} where a 
satisfactory uniqueness result for static region in FLRW was obtained.

The setup consists on a spacetime $(\mmm,g)$ 
composed by two regions $(\mm2^{\stc}, g^{\stc})$ and $(\mm2^{\FL},g^{\FL})$
matched 
in absence of surface layers across their boundaries, denoted by
$\sup^\stc$ and $\sup^\FL$ respectively, which once identified
conform to a hypersurface $\sup$ in $(\mmm,g)$.
The region $\str$ is strictly static, i.e. admits a Killing vector
$\xi$ which is timelike and orthogonal to hypersurfaces everywhere.
$\flr$ is a codimension-zero submanifold with smooth boundary $\sup^{\FL}$ of the FLRW
spacetime
$(\mmm^{\FL},g^{\FL})$, by which we mean the
manifold $\mmm^{\FL} = I\times \M$, where 
$I \subset \mathbb{R}$ is an open interval, $\M$ is either $\mathbb{E}^3$ ($\epsilon=0$), $\mathbb{S}^3$ ($\epsilon=1$) or $\mathbb{H}^3$ ($\epsilon=-1$)
 and the  FLRW metric  $g^{\FL}$ takes the form (\ref{FLmetric}). We call
any coordinate system $\{ R, \theta, \phi \}$ in which the metric takes this
form  a {\it spherical coordinate system}. 
Note that since $(\M,g_{\M})$ is homogeneous, there exist
spherical coordinate systems centered at any point $p \in \M$,
and this will be used below.

The function $a(t)$ is positive and smooth (in fact $C^3$ suffices). 
We assume that $\dot{a}$ does not vanish on any open set (this excludes
uninteresting situations where the FRLW does not evolve). Define the
``geometric''
energy-density $\rho^{\FL}$ and pressure $p^{\FL}$ by
\begin{equation}
\label{eq:def_rhofl}
8 \pi \rhofl  \defi 3  (\dot{a}^2 + \epsilon )/a^2, \quad \quad \quad 
8 \pi \prefl  \defi   (-2 a \ddot{a} - \dot{a}^2 - \epsilon)/a^2,
\end{equation}
so that, if the spacetime has a cosmological constant $\Lambda$, 
the energy-density and pressure of the cosmic fluid 
is  $\rho =  \rho^{\FL} - \frac{\Lambda}{8 \pi} $,
$p = p^{\FL} + \frac{\Lambda}{8 \pi}$. We make the assumption that 
$\rho^{\FL} + p^{\FL} \neq 0$ so that we do have a non-trivial cosmic fluid (this
allows us to define $t$ unambiguously).
Concerning 
the boundary $\sup^{\FL}$ it is assumed to be connected (this is irrelevant
because the matching conditions are local, the assumption is made
merely for notational convenience), 
and nowhere
tangential to a hypersurface of constant cosmic time $t$. This assumption
is physically reasonable and automatically satisfied if the boundary is
causal. In fact, dropping this assumption would only make the
presentation more involved, but would not spoil any of the results
(see \cite{MarsES2} for a discussion).
Finally, we assume that $\sup^{\FL}$ is spatially compact. 
A sufficient condition for this is the 
 ``energy condition'' $\rhofl\geq 0$, see \cite{MarsES1},
and this is in fact the assumption made in \cite{MarsES1}.
However, it can be
proved that spatial compactness suffices for the validity of all the results
below. Note finally, that spatial compactness is indeed an assumption:
allowing for non-compact boundaries, additional configurations not covered
by the uniqueness results do arise, as shown in \cite{MenaNatarioTod}
(see also references therein),
where configurations with planar and hyperbolic symmetries were found
and analized.
It would be interesting to analyze how far can one extend uniqueness without
any compactness assumption on the boundary.
Nevertheless, for the purposes of the Einstein-Straus problem,
compactness is a completely natural assumption, as we want the local
physics unaffected by the cosmological expansion be spatially confined.

By a detailed analysis of the matching conditions,
the following restrictions on the boundary $\sup^{\FL}$ are obtained 
\cite{MarsES2}. First of all, the intersection $S^{\FL}_t$
of $\sup^{\FL}$ with a hypersurface of constant
cosmic time $t$ is a sphere. More precisely,
for any $t \in I$, there exists
a point $c(t) \in \M$ so that $S^{\FL} _t$ is a coordinate sphere of radius
$R(t)$ in a spherical coordinate system centered at $c(t)$.
The radius $R(t)$
is restricted to satisfy the bound
\begin{eqnarray}
\Sigma'^2 - \Sigma^2 \dot{a}^2 |_{R= R(t)} > 0.
\label{Untrapped}
\end{eqnarray}
This inequality means that the surface 
$S^{\FL}_t$ is non-trapped (i.e. has a mean curvature vector spacelike
everywhere). The necessity of this condition can be understood from the fact 
that no closed spacelike surface in a static 
spacetime can have a future (or past) causal and not-identically zero
 mean curvature vector  \cite{MarsSenovillaTrapped}.
The spherical symmetry of the surface $S^{\FL}_t$ and the fact that
the matching conditions force the mean 
curvature vector to be continuous across the matching hypersurface
implies that $S^{\FL}_{t}$ must be non-trapped, which is precisely 
 (\ref{Untrapped}).
Note also that if the static Killing vector $\xi$  admits Killing horizons
and hence changes causal character, then the bound (\ref{Untrapped}) is no
longer necessary.  This behaviour occurs in the Einstein-Straus model
when $a(t)$ is sufficiently small so that $\sigo a(t) = 2m$. 
The breakdown of the ODE (\ref{matchingODE}) in the Einstein-Straus model
is just a manifestation
of the breakdown of the the static coordinate system in the interior
region. In Kruskal
coordinates, the matching would continue across $\sigo a(t) = 2m$
without problem. Something similar
would occur in the general setting if we allowed the static Killing to change
causal character.
 
Returning to the shape of $\sup^{\FL}$, the center point $c(t)$ follows
a geodesic in $(\M,g_{\M})$ (which may degenerate to a point).
The parameter $t$ is not in general an affine
parameter for the geodesic, so the center $c(t)$  is a priori 
allowed to move at any 
speed (in fact the trajectory can have turning points along the
curve). In order to describe the matching hypersurface,
choose any
point $c_0$ along this geodesic and let $c'_0$ be its tangent vector.
We can choose a spherical coordinate system $\{ \hat{R},
\hat{\theta}, \hat{\phi} \}$ centered
at $c_0$ so that the axis of symmetry $\hat{\theta} =0$ is along
the tangent vector $c'_0$.   In this coordinate system,  the center
$c(t)$ will have coordinates
\begin{eqnarray*}
c(t) = \{  \hat{R}= \sigma(t), |\cos \hat{\theta}|  = 1 \}
\end{eqnarray*}
(the value of $\hat{\theta}$ can be $0$ or $\pi$ depending on whether $c(t)$
lies after or before $c_0$ along the geodesic). The function 
$\sigma(t)$ describes the motion of $c(t)$ along the curve. 
The relationship between the spherical coordinate system
$\{ \hat{R},\hat{\theta}, \hat{\phi} \}$ centered at $c_0$ 
and a spherical coordinate system
$\{R,\theta,\phi\}$ centered
at $c(t)$ with parallel axis  (i.e. with coincident lines
$|\cos \theta | = 1$ and $|\cos \hat\theta|=1$) is easily found to be
\begin{eqnarray}
\left .
\begin{array}{rl}
\Sigma(\hat{R}) \sin \hat{\theta} & = \Sigma(R) \sin \theta \\
\Sigma(\hat{R}) \cos \hat{\theta} & = 
\Sigma' (\sigma(t)) \Sigma(R) \cos \theta
+ \Sigma(\sigma(t))  \Sigma' (R )
\\
\hat{\phi} & = \phi
\end{array}
\right \}. \label{coorTrans}
\end{eqnarray}
Thus, the matching hypersurface $\sup^{\FL}$ 
in the spherical coordinates centered at $c_0$ can be parametrized 
by coordinates $t, \theta, \phi$ as
\begin{eqnarray*}
\left . 
\begin{array}{rl}
\Sigma(\hat{R}(t,\theta)) & = \sqrt{\sin^2 \theta \,\, \Sigma^2|_{R(t)} 
+ \left  ( \cos \theta \,\, \Sigma' |_{\sigma(t)} \Sigma |_{R(t)}  
+ \Sigma |_{\sigma(t)}  \Sigma' |_{R(t)}  \right ) ^2} \\ 
\mbox{cotan} (\hat{\theta}(t,\theta))  & = 
\mbox{cotan} (\theta) \,\, \Sigma' |_{\sigma(t)} 
 + \frac{\Sigma |_{\sigma(t)} \Sigma' |_{R(t)}}{\sin \theta \,\, 
\Sigma |_{R(t)}} \\
\hat{\phi} &= \phi
\end{array} \right \}, 
\end{eqnarray*}
which is a rather complicated form for the matching hypersurface.
A useful alternative is to 
use a coordinate system  $\{ t, R, \theta, \phi\}$ in $\flr$ 
such that, for each $t$, $\{ R,\theta,\phi\}$ is the spherical
coordinate system centered at $c(t)$. Applying the coordinate transformation
(\ref{coorTrans}) to the metric
\[
g^{\FL} =  -dt^2 + a^2(t) \left ( 
d \hat{R}^2 + \Sigma^2(\hat{R}) \left ( d \hat{\theta}^2 +
\sin^2 \hat{\theta} d \hat{\phi}^2 \right ) \right )
\]
gives
\begin{align}
ds^2 = & -dt^2 + a^2 (t) \left [ \left (dR + f(t) \cos \theta dt \right )^2 
+ \left ( \Sigma(R) d\theta  - f(t) \Sigma' (R)
\sin \theta dt \right )^2 + \right . \nonumber  \\
& + \left .   \frac{}{}
\Sigma^2 (R) \sin^2 \theta d\phi^2 \right ].
\label{tiltedForm}
\end{align}
where $f(t)  = \dot{\sigma}(t)$. The explicit calculation leading to
(\ref{tiltedForm}) is somewhat long,
a much more elegant method of obtaining this form
of the metric is discussed in the Appendix of \cite{MarsES2}. In these
coordinates, the matching hypersurface $\sup^{\FL}$ is simply
$\{ R = R(t) \}$.

The matching conditions in the FLRW part are supplemented with 
a differential equation relating the trajectory of the
center $\sigma(t)$ with the radius $R(t)$. To write it down,
define a function $\beta(t)$ via
\begin{equation}
\tanh \beta(t) = \left . \epsilon_1
\frac{\Sigma \dot{a}}{\Sigma'} \right |_{R=R(t)},
\label{tanhbeta}
\end{equation}
where $\epsilon_1 = \pm 1$ depending on whether the FLRW
region to be matched is $R> R(t)$ or $R< R(t)$ (more specifically,
the manifold with boundary $\mm2^{\FL}$ is $\{ \epsilon_1 R \geq 
\epsilon_1 R(t) \}$).  Note that $\beta(t)$ is well-defined because of
(\ref{Untrapped}). The ODE relating $R(t)$ and $\sigma(t)$
is conveniently written using an auxiliary function $\Delta(t) \neq 0$
as the following pair of differential equations
\begin{align}
& \dot{f} = 
\left ( \frac{2 \epsilon_1 \dot{a} \dot{R}}{\tanh \beta} + 
\frac{\dot{\Delta}}{\Delta} - \frac{2 \cosh \beta}{\sinh \beta}
\dot{\beta} \right )f,  \nonumber \\
& f^2 \left . \frac{\Sigma'}{\Sigma} 
\right |_{R=R(t)} 
-\ddot{R} + \frac{\epsilon_1 \dot{a} \dot{R}^2}{\tanh \beta} +
\frac{\dot{\Delta}}{\Delta} \left ( \dot{R} + 
\frac{\epsilon_1}{a \tanh \beta}
\right ) - 2  \left ( \frac{\epsilon_1}{a} + 
\frac{\dot{R}}{\tanh \beta} \right )  \dot{\beta}
 =0. \label{RelEqs}
\end{align}
In summary, the static
domain embedded in the FLRW regions consists of a sphere with time
dependent radius $R(t)$
and with center moving across the FLRW spacetime along a geodesic.
Speed along the geodesic and radius $R(t)$ 
are linked by (\ref{RelEqs}). The model is not
spherically symmetric because the center of the sphere is allowed to move.
However, it
is  very close to spherically symmetric and  it turns out that 
the center of the sphere must be at rest for several relevant matter models
in the static domain, as we discuss next.

An important consequence 
of the matching procedure (cf. Lemma 1 in \cite{MarsES2}) is that the
Killing vector field $\xi$ is everywhere transverse to
the matching hypersurface $\sup^{\stc}$ in the static region. 
Combining this with the fact that the static geometry is invariant along 
$\xi$ it follows that the static metric 
in the spacetime region obtained by dragging $\sup^{\stc}$ with
the static Killing vector can be
fully determined in terms of  hypersurface geometry of $\sup^{\stc}$.
Since, in turn, the geometry
on $\sup^{\stc}$ is related to the geometry of $\sup^{\FL}$
via the matching conditions, it follows that the spacetime
geometry of the static region
becomes completely 
determined \cite{MarsES2} in a neighbourhood of its matching hypersurface
in terms of the FLRW geometry and the functions $R(t), \sigma(t)$
and $\Delta(t)$ (see Theorem 1 in \cite{MarsES2} for  details).
Specifically, there exist coordinates $\{ T, t, 
\theta, \phi\}$ so that the metric $g^{\stc}$ takes the form (note that $t$
is a spacelike coordinate in the static domain)
\begin{eqnarray}
g^{\stc} = - \frac{ (\cosh \beta + \mu \sinh \beta )^2 }{\Delta^2} dT^2
+ \left (\mu \cosh \beta + \sinh \beta  \right )^2 dt^2 
+ \hspace{2cm} \nonumber \\
+ 
a^2(t) \Sigma^2(R(t))
\left [ \left (d\theta - f(t) \left .
\frac{\Sigma'}{\Sigma} \right |_{R=R(t)} \sin \theta dt \right )^2
+ \sin^2 \theta d\phi^2 \right ],
\label{staticmetric}
\end{eqnarray}
where $\mu \defi \epsilon a(t) ( \dot{R}(t) + f(t) \cos \theta)$.
The matching hypersurface  $\sup^{\stc}$ is defined by the embedding 
$\{t,\theta,\phi \} \rightarrow \{T=T(t), t, \theta, \phi\}$,
where $T(t)$ satisfies $\dot{T}(t) = \Delta(t)$
and the portion of the static spacetime to be matched to the exterior
region is $\{ T \geq T(t) \}$. The metric (\ref{staticmetric}) is foliated 
by round  spheres $\{ T= \mbox{const}, t = \mbox{const} \}$ but it is not
spherically symmetric in general (unless $f(t)=0$ and $R(t)=0$). 
To complete the picture, we review the 
energy-momentum tensor in the static part. Introduce
two one-forms and one symmetric two-tensor $h$ by
\begin{align*}
\theta^0 &= \frac{ (\cosh \beta + \mu \sinh \beta )}{\Delta} dT, \quad \quad
\theta^1 = \left (\mu \cosh \beta + \sinh \beta  \right ) dt, \\ 
h &= a^2(t) \Sigma^2(R(t))
\left [ \left (d\theta - f(t) \left .
\frac{\Sigma'}{\Sigma} \right |_{R=R(t)} \sin \theta dt \right )^2
+ \sin^2 \theta d\phi^2 \right ],
\end{align*}
The Einstein tensor of $\str$ is (cf. Proposition 1 in \cite{MarsES2})
\[
\frac{1}{8\pi} G^{\stc} = \rho^{\stc} \theta^{0} \otimes  \theta^{0}
+ p_r^{\stc} \theta^{1} \otimes \theta^{1}
+ p_t^{\stc} h 
\]
where $\rho^{\stc}$, $p^{\stc}_r$ and $p^{\stc}_t$ read
\begin{align}
\rho^{\stc} & = \frac{\rho^{\FL} \mu -p^{\FL} \tanh \beta}{\mu + \tanh \beta }, 
\nonumber \\
\rho^{\stc} + p_r^{\stc} & = \frac{\left ( \rho^{\FL} + p^{\FL} \right ) \mu}{
\left ( \mu \sinh \beta  + \cosh \beta \right ) \left ( \sinh \beta 
+ \mu \cosh \beta \right )}, \label{staticEM} \\
p^{\stc}_t & =  \frac{3 p^{\FL} - \rho^{\FL}}{6} + 
\frac{\tanh \beta
\left (2 f^2 \sin^2 \theta 
 - \Sigma^2 (R(t)) \left (\rho + p \right ) \left (\mu^2 -1 \right ) \right ) }
{16 \pi \Sigma^2 (R(t)) \left ( \mu + \tanh \beta \right ) \left (
1 + \mu \tanh \beta\right )} + \hspace{2cm} \nonumber \\
& + \frac{\left [ -\ddot{\beta}+
\frac{\dot{\Delta}}{\Delta} \left (\frac{\dot{a}}{a} \frac{1}{\tanh \beta}
+ \dot{\beta} \right ) -  \frac{2 \dot{\beta} \dot{a}}{a} 
 \left ( 1 + \frac{\mu}{\tanh \beta} \right ) 
+ \mu \left (\frac{\ddot{a}}{a} + \frac{\mu \dot{a}^2}{a^2 \tanh \beta}
\right ) \right ]}{8 \pi \cosh^2 \beta \left (\mu  + \tanh\beta \right ) 
\left (1 + \mu \tanh \beta \right )}, \nonumber 
\end{align}
and $\rho^{\FL}$ and $p^{\FL}$ were defined in (\ref{eq:def_rhofl}). 
Several consequences of (\ref{staticEM}) can be drawn \cite{MarsES1,MarsES2}.
Concerning the uniqueness of the Einstein-Straus model, under the assumption
$\rho^{\stc} + p^{\stc} =0$
(which includes vacuum with or without cosmological constant or
a non-singular electromagnetic field) it follows that $\mu=0$ and hence $f(t)=0$
and $R(t)=0$. The second equation in (\ref{RelEqs})  gives (with an
appropriate but still completely general choice of integration constant),
\begin{eqnarray*}
\dot{T}(t) = \Delta (t) = \frac{1}{\Sigma'_c} \cosh^2 \beta(t) = 
\frac{\Sigma'_c{}}{\Sigma'_c{}^2-\Sigma_c^2 \dot{a}^2},
\end{eqnarray*}
where (\ref{tanhbeta}) and the definitions  of $\Sigma_c$ and $\Sigma'_c$ in 
Section \ref{sec:ESmodel} have been used. The static metric simplifies to
\begin{eqnarray*}
g^{\stc} =   - \left ( 
\Sigma'_c{}^2 - \Sigma_c^2  \dot{a}^2 \right ) dT^2 +  \frac{\Sigma_c^2 \dot{a}^2}{\Sigma'_c{}^2
- \Sigma_c^2
\dot{a}^2} dt^2 + \Sigma_c^2 a^2(t) d \Omega^2.
\end{eqnarray*}
From this metric, uniqueness of the Einstein-Straus model as the unique static
vacuum (with or without cosmological constant) region embedded in a 
FRLW cosmological model follows easily. So, static vacuoles
in a FLRW model must be spherically symmetric both in shape and
interior geometry. 

It is an open problem to analyze whether
there are any physically realistic matter models in the interior
static region for which the motion of the static domain inside FLRW
has interesting properties. Note that a priori nothing prevents the
geodesic $c(t)$ from being spacelike, so the motion of the 
static domain can in principle be superluminal for the cosmic
observers. This is of course reminiscent to  the superluminal warp drive
discovered by Alcubierre \cite{Alcubierre}.

\subsection{Uniqueness results in the stationary and axisymmetric case}
\label{sec:STAX}
The study of axially symmetric equilibrium regions in FLRW universe
was dealt with in \cite{Nolan_Vera_05}, and, in short, the main result found
was that those stationary regions must, in fact, be static, and therefore
the results of the previous section apply.

With the same definitions and assumptions concerning the FLRW region
$(\mm2^{\FL},g^{\FL})$ and its boundary $\sup^{\FL}$
as in the previous section, we assume now that the interior region
$(\mm2^{\sx}, g^{\sx})$ is (strictly) stationary and
axisymmetric. 
More specifically, we demand that
(i) the spacetime admits a two-dimensional group of isometries
$G_2$ acting simply-transitively on timelike surfaces $T_2$
and containing a (spacelike) cyclic subgroup, so that $G_2=\mathbb{R}\times \mathbb{S}^1$,
and (ii) that the set of fixed points of the cyclic group is not empty.
Consequences of the definition are that
the $G_2$ group has to be Abelian
\cite{commu,Carot99,Barnes00},
and that the set of fixed points
must form a timelike two-surface \cite{maseaxconf}, which is the axis.
The axial Killing $\ax$ is then intrinsically defined by
normalizing it demanding
$\partial_\alpha\ax^{\,2}\partial^\alpha\ax^{\,2}/4\ax^{\,2}\to 1$
at the axis.
See also \cite{solutions_book} and \cite{jaumeax}.
In addition, we demand that the isometry group is
orthogonally transitively (OT), i.e. that the two planes orthogonal
to the orbits of the isometry group are surface forming. 
This assumption is also known as the ``circularity condition'', and
in many cases of interest it follows
as a consequence of the
Einstein field equations. Indeed, the $G_2$ on $T_2$ group must act orthogonally
transitively in a region that intersects the axis of symmetry
whenever the Ricci tensor has an invariant 2-plane spanned by the
tangents to the orbits of the $G_2$ on $T_2$ group
\cite{carter69}. By the Einstein field equations, this includes
$\Lambda$-term type matter (i.e. vacuum with or without cosmological
constant), perfect fluids without convective motions, and also
stationary and axisymmetric electrovacuum \cite{solutions_book}.

An OT stationary and axisymmetric
spacetime $\spsx$ is locally characterized by the existence
of a coordinate system $\{T,\tphi,x^M\}$
($M,N,...=2,3$) in which the line-element for the
metric $g^{\sx}$ outside the axis
reads \cite{solutions_book}
\begin{equation}
  \label{eq:ds2sx}
  g^{\sx}=-e^{2U}\left( d T+Ad\tphi\right)^2 +
e^{-2U}W^2 d \tphi^2 + g_{MN}d x^M d x^N,
\end{equation}
where $U$, $A$, $W$ and $g_{MN}$ are functions of $x^M$,
the axial Killing vector field is given by $\ax=\p_{\tphi}$,
and a timelike (future-pointing) Killing vector field is given by
$\stk=\p_{T}$. Although useful for the sake of clarity, the use of 
coordinates is not essential for the results below, which only depend
on the intrinsic geometric properties of of $\sxr$.

As before, no specific matter content
is assumed in the stationary and axisymmetric region.
Regarding the matching hypersurface $\sup$, besides those
in the previous Section \ref{sec:STATIC}, we make the only extra assumption
that it preserves the
axial symmetry \cite{mps} of $\sxr$ and of $\flr$.
This means that $\sup^{\sx}$ 
is assumed to be invariant under the axial symmetry of $\sxr$ and
that there is an axial Killing vector $\axfl$
in the Killing algebra of $\flr$ tangent to 
$\sup^{\FL}$.

With these assumptions at hand, in \cite{Nolan_Vera_05},
it is proven, first, that the stationary (timelike) Killing vector field $\xi$
is nowhere tangent to $\sup^{\sx}$.
As explained in the previous section,
this serves in particular
to construct a neighbourhood
of $\sup^{\sx}$ by dragging it along the orbits of $\xi$,
in which the geometry is thus fully determined by the information in $\sup^{\sx}$.
The main result in \cite{Nolan_Vera_05} is that
if a OT stationary and axisymmetric region $\sxr$
can be matched to a FLRW region through a hypersurface
$\sup^{\sx}$ preserving the axial symmetry,
then the region $\sxr$ must be, in fact, static
on a neighbourhood 
of $\sup^{\sx}$. 

All in all, in that neighbourhood of the matching hypersurface
the OT stationary axisymmetric region $\sxr$
thus becomes a static axisymmetric region $\str$, and therefore the
results of the previous Section \ref{sec:STATIC} (cf. \cite{MarsES2}) apply.

So far, no explicit condition on the matter content of the stationary region
has been used. When conditions on the matter content on the
(OT) stationary and axisymmetric (and hence static) region are imposed,
the functions that determine the matching hypersurface
and the static geometry are determined
by the results reviewed in Section \ref{sec:STATIC}.

In particular, and for completeness, let us consider a vacuum
(with or without a cosmological constant)
stationary and axisymmetric region $\sxr$ matched to FLRW
preserving the axial symmetry.
As mentioned above, a \emph{vacuum} matter content forces
the axial symmetry and stationary group $G_2$ on $T_2$ to act
orthogonal transitively.
On the other hand, a region of FLRW $(\mm2^{\FL},g^{\FL})$
matched to a vacuum region must satisfy the assumptions
made and, in particular, have a causal $\sup^\FL$ (in fact
tangent to the fluid flow).
The above result implies then that the vacuum region
must be static. The results
of Section \ref{sec:STATIC} thus apply, and 
imply, in turn, that the whole region (not just its boundary)
has to be spherically symmetric,
and hence Schwarzschild.

To sum up, this means that the only stationary and axially symmetric vacuum
region that can be matched to FLRW is a spherically symmetric piece of
Schwarzschild. This constitutes still another uniqueness result of
the Einstein-Straus model when the vacuum region lies inside
FLRW.

The complementary result, when the vacuum region lies outside FLRW,
constitutes a uniqueness result of the Oppenheimer-Snyder model \cite{OPSN}.
It is worth noticing that this result can also be interpreted
as a no-go result  for the possible interiors of Kerr. Indeed, this
result states that an axially symmetric region of (an evolving) FLRW,
irrespective of its relative rotation with the exterior,
cannot be the source of a stationary and axisymmetric vacuum region,
in particular, Kerr.

\section{Robustness of Einstein-Straus: Generalized exact cosmologies}
\label{sec:robustness}

In the previous sections, the robustness of the Einstein-Straus model has been discussed
by considering generalizations of the interior vacuole and keeping the
FLRW model as the exterior cosmological model. The question arises as
to what happens if these symmetry assumptions concerning the exterior
metric are relaxed. There are two different ways to study
departures from the spherical FLRW: either perturbatively or using exact
solutions. 
In this section, we concentrate on the later possibility
and, in particular, we review the results found in \cite{MTV} for spatially homogeneous (but anisotropic) cosmologies.

A step in the direction of generalising the exterior was taken by Bonnor, who considered the
embedding of a Schwarzschild region in an expanding spherically
symmetric inhomogeneous Lema\^itre-Tolman-Bondi (LTB) 
exterior. He found that
such a matching is possible in general, and that it allows the average
density of the Schwarzschild interior to be chosen independently of
the exterior LTB density \cite{Bonnor00}. So, clearly, if the spherical
symmetry is kept, the exterior can be readily generalized to the case
of inhomogeneous dust cosmologies.  An interesting review about physical aspects of
spherically symmetric Einstein-Straus and McVittie type models 
can be found in \cite{Carrera}. 

While keeping the spherical symmetry of the cavity allows for
straightforward generalizations of Einstein-Straus, breaking this symmetry brings
in unexpected complications, as we have seen in the previous section. The same seems to
hold if we consider exact generalizations of the FLRW cylindrical
symmetry exterior to spatially homogeneous but anisotropic
spacetimes. This problem was considered in \cite{MTV} and we
summarize it next:

Consider the problem of matching, across a hypersurface $\sup$ which is spatially a topological
sphere\footnote{The assumption made in \cite{MTV} is that
the region $(\mm2^{\stc},g^{\stc})$ was simply connected, in order
to force $\sup^{\stc}$ being spatially a topological sphere.
We prefer to impose here the assumption directly on $\sup^{\stc}$
and leave the possibility of having more general interiors.}, 
of an interior locally cylindrically
symmetric (LCS) static spacetime $(\mm2^{\stc},g^{\stc})$, which
represents a spatially compact 
region, to a
spatially homogeneous anisotropic exterior $(\mm2^{\HOM},g^{\HOM})$
having a Lie group $G_3$ acting on $S_3$ surfaces.
Once more, we make no
assumptions {\em a priori} on the matter contents,
so the main results are purely geometric. Only at the end of the section, assumptions on the matter content will lead to a final no-go result.

We wish
to preserve globally the axial symmetry and thus we need to impose first
the existence of a group of cyclic symmetries in the exterior region.
This, combined
with the assumption of a spacelike spherical topological
shaped\footnote{This implies the existence of north and south poles
on $\sup^{\stc}$
where the axial killing vector in the interior vanishes
and, therefore, also the generator of the cyclic
symmetry in the exterior region on  $\sup^{\HOM}$ by construction.}
$\sup^{\stc}$, 
implies
the existence of an exterior axis.
Thus, the cyclic
symmetry in the exterior must really be an axial symmetry (see the discussion in Section \ref{sec:STAX}). Moreover, we require that the cylindrical 
symmetry is preserved (in the sense of \cite{mps})
at least on a non-empty open
subset of $\sup$.
This has the consequence that
$(\mm2^{\HOM},g^{\HOM})$ must admit a $G_4$ on $S_3$ group of isometries,
so that it is locally rotationally symmetric (LRS).
In \cite{MTV}, it is shown how all the LRS
spatially homogeneous metrics can be written 
in the form adapted to the two commuting Killing vectors defining
the cylindrical symmetry $\partial_\varphi$
and $\partial_z$, where $\partial_\varphi$ is axial with axis at $r=0$:
\begin{equation}
\label{non-static}
g^{\HOM}=-dt^2+ b^2(t)dr^2-2\epsilon r b^2(t) dr dz+\hat{C}^2
d\varphi^2+2\hat{E} dz d\varphi+\hat{D}^2d{z}^2
\end{equation}
with
\begin{eqnarray}
&&\hat{C}^2=b^2(t)\Sigma^2(r,k)+na^2(t)(F(r,k)+k)^2,\nonumber\\
&&\hat D^2 = a^2(t) + \epsilon r^2 b^2(t),~~~\hat E= na^2 (t)(F(r,k)+k),\label{coeficients}
\end{eqnarray}
where the functions $\Sigma$ and $F$ are given by
\[
\label{sigma}
\Sigma(r,k)=
\left\{
\begin{array}{cl}
\sin r,& k=+1\\
r, & k=0\\
\sinh r, & k=-1
\end{array}
\right.
~~
~~
\text{and}
~~
F(r,k)=
\left\{
\begin{array}{cl}
-\cos r,& k=+1\\
r^2/2, & k=0\\
\cosh r, & k=-1,
\end{array}
\right.
\]
and where $\epsilon$ and $n$ are given such that
$\epsilon=0,1;~~n=0,1;~~\epsilon n=\epsilon k=0$\footnote{Let
us note that the case $\epsilon=n=0$ with $k=1$ is special, since it
corresponds to the Kantowski-Sachs (KS) class of metrics, which do not
admit a $G_3$ on $S_3$ subgroup \cite{solutions_book}.
It was included in the study for completeness.}.
The metrics are classified according to these constants in Table \ref{table:bianchis}.  
\begin{table}[h!]
\begin{center}
\begin{tabular} {c c  c  c }
Bianchi types & $\epsilon$&$n$&$k$ \\
\hline
I, VII$_0$ &0&0&0
\\
III & 0&0&-1
\\
IX & 0&1&1
\\
II & 0&1&0
\\
VIII,III &0&1&-1
\\
V,VII$_h$ &1&0&0
\\
\hline
\end{tabular}
\caption{\label{table:bianchis} Classification of possible
$G_3$ on $S_3$ subgroup types according to the values
of $\{\epsilon,k,n\}$ for the metric given by (\ref{non-static}).
}
\end{center}
\end{table}

The matching conditions are then investigated and a crucial step was to observe that they lead to the following necessary relations involving only exterior metric functions:
\begin{equation}
\label{exterior}
\hat{D}_{,t}\hat{C}_{,r}-\hat{D}_{,r}\hat{C}_{,t}\stackrel{\Omega}{=}0,~~~
\hat{E}_{,t}\hat{D}_{,r}-\hat{E}_{,r}\hat{D}_{,t}\stackrel{\Omega}{=}0,~~~
\hat{E}_{,t}\hat{C}_{,r}-\hat{E}_{,r}\hat{C}_{,t}\stackrel{\Omega}{=}0,
\end{equation}
which, for non-static exteriors,
imply $n=0$ and thus exclude Bianchi types II, III, VIII and IX.

By inserting (\ref{exterior}) back in the matching conditions one is able to prove a series of results that lead to the following conclusion: The only expanding spatially homogeneous spacetimes
which can be matched to a 
locally cylindrically symmetric static interior region preserving
the (cylindrical) symmetry,
across a non-spacelike hypersurface which is
spatially a topological sphere,  are given by
\begin{equation}
\label{tiran}
ds^2=-dt^2+\beta^2d z^2+b^2(t)\left[(dr-\epsilon rdz)^2+\Sigma^2(r,k) d\varphi^2\right],
\end{equation}
where $\beta$ is constant. This metric for $k=1$ belongs to the Kantowski-Sachs class, when $k=-1$ admits a $G_3$ on $S_3$
of Bianchi type III, and when $k=0$ of Bianchi types I,V,VII${}_0$,VII${}_h$.

The metrics (\ref{tiran}) are very special, partly due to the fact that 
condition $a_{,t}=0$ (see (\ref{non-static})-(\ref{coeficients})) imposes a strong constraint, implying that there
cannot be any time (nor space) evolution along the direction orthogonal to the orbits of the subgroup
$G_3$ on $S_2$ of the LRS. There are also constraints imposed through the matching in the interior region and the interested reader can find those 
in \cite{MTV}. Note that the no-go result found in \cite{Seno_Vera_97} is trivially recovered, since
FLRW is included in the LRS class for $b(t)=a(t)$, and the above
would imply a static FLRW metric.

If one specifies a particular class of matter fields, then one gets further constraints on the cosmological dynamics. For example, if the matter in the static region is not specified but the dynamical (cosmological) region is assumed to contain a perfect fluid with pressure $p$ and energy-density $\rho$ satisfying the dominant energy condition everywhere, then $\epsilon=0$ necessarily, which corresponds to a stiff fluid equation of state
$\rho=p=\alpha^2/(4t^2(\alpha-kt)^2)$,
with  $b(t)=\sqrt{\alpha t-kt^2}$,
where $\alpha>0$. On the other hand, if the interior is vacuum then the exterior must be also vacuum.

The overall no-go results, in this case, can be seen in two ways: either
as a consequence of the assumption that the 
interior metric is static and cylindrically symmetric, which 
seems to prohibit time dependence along one direction, 
or as a consequence of
the particular exterior metrics we are considering,
which are homogeneous then prohibiting the coefficients along this direction
to be space dependent.
The perturbative approach described in the next sections
can help to clarify this question.

To conclude this section we remark that there exists an 
example of a Einstein-Straus model with
exact non-spherical inhomogeneous cosmologies. This is provided by the Szekeres dust solution which has no Killing vectors, in general, but contains intrinsic symmetries on 2-spaces of constant curvature:
The Szekeres solution is divided into class I, which generalizes the LTB solution having non-concentric spheres of constant mass, and class II which includes the Kantowski-Sachs solution. 
Class I solutions have been proved to be interiors to
the Schwarzschild solution \cite{Bonnor76} and this result has been
generalized to include the cosmological constant \cite{Lake00,
MenaNatarioTod}.
As mentioned before, one can invert the
roles of the two spacetimes involved and hence
construct an Einstein-Straus type model with a Schwarzschild or Kottler cavity
within a class I Szekeres' cosmology. 
Let us remark that the Szekeres class has been used recently
in Swiss cheese models as interiors
to FLRW (see \cite{Bolejko09} and references therein).

Class II Szekeres dust metrics are less known, but contain curious inhomogeneous solutions with cylindrical symmetry
\cite{Seno_Vera00}. In this case, it seems harder to be able to get a
physically reasonable Einstein-Straus model considering what we have described above.

\section{Brief overview of perturbative matching theory}
\label{sec:overview}
A perturbed spacetime consists of a symmetric two-covariant 
tensor (the ``perturbation metric'') defined on a fixed spacetime
(the ``background''). From a structural point of view, spacetime perturbation
theory is a gauge theory in the sense that many perturbation metrics
describe the same physical situation (i.e. they are gauge related). The underlying
geometrical reason for this gauge freedom can be understood from 
the following intuitive picture of perturbation theory. We imagine
a one-parameter family of spacetimes $(\mmm_{\varepsilon},g_\varepsilon)$ such 
that all the manifolds are diffeomorphic to each other.
This allows one to pull back $g_{\varepsilon}$  onto a single 
manifold in the family (say $\mmm_{0} \defi \mmm_{\varepsilon=0}$)
and work with a one-parameter family of metrics on a single manifold.
If all the construction is smooth in $\varepsilon$, derivatives with respect
to this parameter can be taken. The perturbation metric $\gfpert$
is simply the derivative
at $\varepsilon=0$ of this family of metrics. However, the identification
of points in the different manifolds (the diffeomorphism above) is highly
non-unique. Any other choice of identification would lead to a different,
but geometrically equivalent, 
perturbation metric. This is
the gauge freedom of the theory. Intuitively, it is clear that the gauge freedom
will consist of a vector field on the background, because this measures
the shift of the new identification with respect to the previous one, and
an initial direction (in $\varepsilon$) is all what is required to compute
derivatives with respect to $\varepsilon$ at $\varepsilon=0$. 

When two spacetimes with boundary are matched, an identification of 
the boundaries is required. As already mentioned before,
if the boundaries are nowhere null,
the matching conditions require the  equality of the induced metric and
second fundamental form (with appropriate choices of orientation).
To compare the tensors it is necessary to pull them back to a single
manifold and this is done via the identification of boundaries.
Contrarily than before, the matching theory is strongly dependent
on the identification of the boundaries.  In fact,  the matching conditions
demand the {\it existence} of one such identification for which 
the first and second fundamental forms agree. 

Assume now that we are studying perturbation theory on a
background spacetime constructed from
the matching of two spacetimes.
The question then arises of what are the conditions that the 
metric perturbation tensors on each side must satisfy to
have a perturbed matching spacetime.  This issue is somewhat more
involved than one may think a priori and it was solved in a complete manner
for the first time by Battye and Carter \cite{BattyeCarter01}
and independently
by Mukohyama \cite{Mukohyama00} (this has been
 extended to second order in \cite{Mars05}).
Previous attempts \cite{GS_junction_odd_79,GS_junction_even_79,Garcia_Gundlach_matching_01} did not take into
account all the subtleties of the interplay between two completely
different gauge freedoms inherent to  this problem. Indeed,
in the picture above of perturbation theory in 
terms of a collection of 
spacetimes $(\mmm_{\varepsilon},g_{\varepsilon})$, each one of them
arises now as the matching of two spacetimes with boundary
$\mm2^{\pm}_{\varepsilon}$ across their respective
boundaries $\sup^{\pm}_{\varepsilon}$. For better visualization, 
assume that each one of $\mm2^{\pm}_{\varepsilon}$ is a submanifold with boundary
of a larger boundary-less manifold  $\mmextended^{\pm}_{\varepsilon}$ and
assume that the $\{ \mmextended^{+}_{\varepsilon}  \}$ manifolds are identified among themselves
(say with $\mmextended^+_0 \defi \mmextended^+_{\varepsilon=0}$) via an $\varepsilon-$dependent diffeomorphism. The hypersurface $\sup^{+}_{\varepsilon}$ projects down to $\mmextended^+_0$
as a hypersurface $\widehat{\sup}^+_{\varepsilon}$. Now we have a collection
of hypersurfaces in one single manifold, and one can think of
taking $\varepsilon-$derivatives of geometric quantities intrinsic to 
the hypersurface. The important point is that, given a point $p\in
\sup^+_{\varepsilon=0}$, we do not know how this point maps into 
$\widehat{\sup}^+_{\varepsilon}$. For that, it is necessary to prescribe 
{\it first} how $p$ is mapped into $\sup^+_{\varepsilon}$. The identification of
$\{ \sup^+_{\varepsilon} \}$ among themselves is an additional gauge freedom.
It is fully independent of the standard 
gauge freedom in perturbation theory 
(called ``spacetime gauge freedom'' from  now on) 
and is referred to as
{\it hypersurface gauge freedom} \cite{Mukohyama00}. The composition of both 
identifications gives, as $\varepsilon$ varies, and for any $p
\in \sup^{+}_{\varepsilon=0}$, a path $\gamma_p(\varepsilon)$ in $\mmextended^+_0$ starting
at $p$. Since everything is smooth in $\varepsilon$, the tangent vector
to this path at $\varepsilon=0$ defines a vector field  $Z^+$ on
$\sup^+_0  \defi \sup^+_{\varepsilon=0} (= \widehat{\sup}_{\varepsilon=0})$.
This vector field is not necessarily tangent (nor normal) 
to $\sup^+_0$ and it depends on both gauge freedoms. A schematic
figure for the definition of $Z^+$ and how it depends on the gauges is
given in Figure \ref{fig:vecZ}. If we let $n_+^{(0)}$ be a unit normal
vector to $\sup^+_0$, we can decompose $Z^+ = Q^+ n_{+}^{(0)} + T^+$,
where $T^+$ is tangent to $\sup^+_0$. From the discussion above,
it should be clear that $Q^+$ is independent of the hypersurface
gauge while  $T^+$ strongly depends on it.
In fact, it can always be made zero by an appropriate choice of gauge.
However, doing this is not usually a good idea because the matching
has two regions and, at each value of $\varepsilon$, the matching requires
\label{fig:vecZ}
an identification between $\sup^+_{\varepsilon}$ and 
$\sup^-_{\varepsilon}$. After a choice of hypersurface
gauge to identify
$\sup^+_{\varepsilon}$ with $\sup^+_{0}$ we have no freedom left
to choose a hypersurface gauge to identity $\sup^-_{\varepsilon}$
and $\sup^-_{0}$. The matching conditions will tell us
how this identification must be done. So, had we chosen
$T^+=0$, we would still have to leave $T^-$ free and let
the linearized matching theory determine its value.
Further details on the double gauge freedom of linearized
perturbation theory can be found in the paper of Mukohyama 
\cite{Mukohyama00}
and in \cite{ESpert1},
where the issue is discussed in depth including a critical
analysis of previous attempts to formulate a consistent
perturbative matching theory.

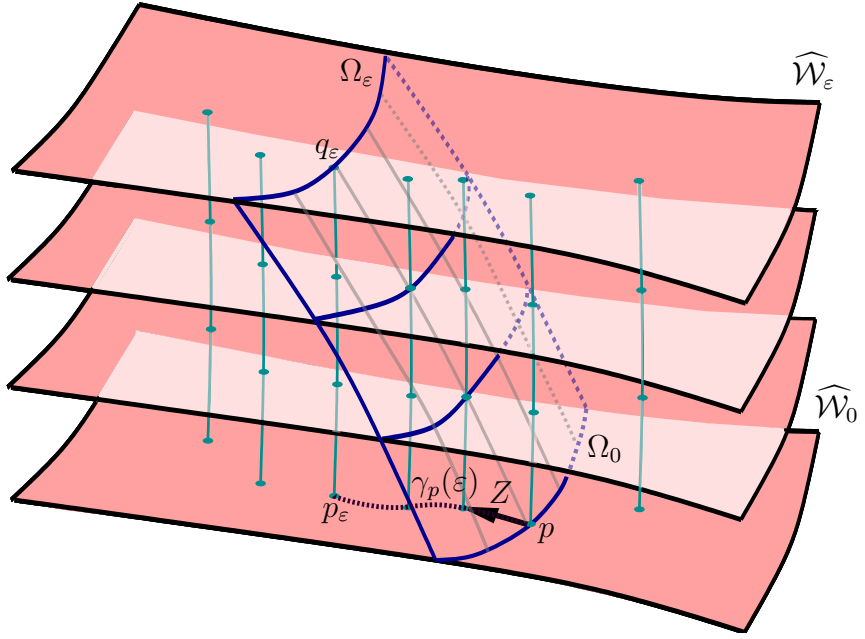
\begin{figure}[h]
  \centering
  \newcommand{\svgwidth}{15cm}
\begingroup%
  \makeatletter%
  \providecommand\color[2][]{%
    \errmessage{(Inkscape) Color is used for the text in Inkscape, but the package 'color.sty' is not loaded}%
    \renewcommand\color[2][]{}%
  }%
  \providecommand\transparent[1]{%
    \errmessage{(Inkscape) Transparency is used (non-zero) for the text in Inkscape, but the package 'transparent.sty' is not loaded}%
    \renewcommand\transparent[1]{}%
  }%
  \providecommand\rotatebox[2]{#2}%
  \ifx\svgwidth\undefined%
    \setlength{\unitlength}{655.20004883bp}%
    \ifx\svgscale\undefined%
      \relax%
    \else%
      \setlength{\unitlength}{\unitlength * \real{\svgscale}}%
    \fi%
  \else%
    \setlength{\unitlength}{\svgwidth}%
  \fi%
  \global\let\svgwidth\undefined%
  \global\let\svgscale\undefined%
  \makeatother%
  \begin{picture}(1,0.56043952)%
    \put(0,0){\includegraphics[width=\unitlength]{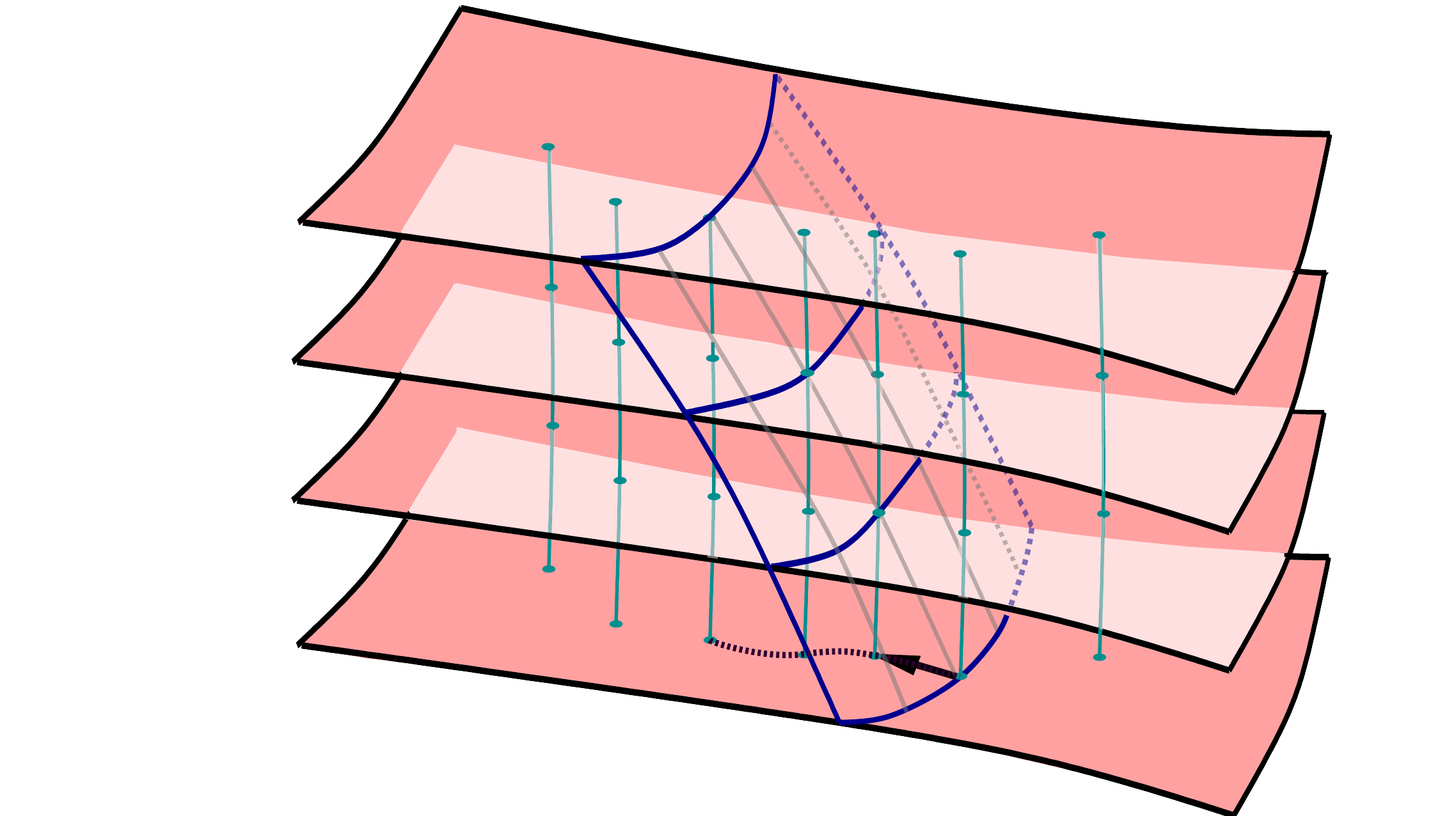}}%
    \put(0.55559827,0.12467073){\color[rgb]{0,0,0}\makebox(0,0)[lb]{\smash{$\gamma_p(\varepsilon)$}}}%
    \put(0.88932283,0.48437499){\color[rgb]{0,0,0}\makebox(0,0)[lb]{\smash{$\mmextended_\varepsilon$}}}%
    \put(0.90926194,0.19024007){\color[rgb]{0,0,0}\makebox(0,0)[lb]{\smash{$\mmextended_0$}}}%
    \put(0.49293668,0.4872951){\color[rgb]{0,0,0}\makebox(0,0)[lb]{\smash{$\sup_\varepsilon$}}}%
    \put(0.70975703,0.15508383){\color[rgb]{0,0,0}\makebox(0,0)[lb]{\smash{$\supo$}}}%
    \put(0.62340672,0.11466462){\color[rgb]{0,0,0}\makebox(0,0)[lb]{\smash{$Z$}}}%
    \put(0.47091932,0.42372926){\color[rgb]{0,0,0}\makebox(0,0)[lb]{\smash{$q_\varepsilon$}}}%
    \put(0.4776791,0.10217698){\color[rgb]{0,0,0}\makebox(0,0)[lb]{\smash{$p_\varepsilon$}}}%
    \put(0.66710188,0.08417585){\color[rgb]{0,0,0}\makebox(0,0)[lb]{\smash{$p$}}}%
  \end{picture}%
\endgroup%
  \caption{
The different spacetimes $\mmextended_\varepsilon$ are represented by horizontal
sheets, with $\mmextended_0$ at the bottom. On each
$\mmextended_\varepsilon$ there is a hypersurface $\sup_\varepsilon$.
The hypersurfaces $\sup_\varepsilon$ for all $\varepsilon$
span a manifold transverse to $\mmm_\varepsilon$ (the blue sheet in the figure).
The choice of spacetime gauge is represented by big dots
on each $\mmextended_\varepsilon$ linked by green curves, while the choice
of hypersurface gauge is represented by the grey curves on the
blue sheet. The point $p$ is mapped first to $q_\varepsilon$ through the
hypersurface gauge and back to $p_\varepsilon\in \mmextended_0$ through the
spacetime gauge. The points $p_\varepsilon$ define on $\mmextended_0$
the path $\gamma_p(\varepsilon)$. The vector $Z$ is the tangent of
$\gamma_p(\varepsilon)$ at $p$.} 
\end{figure}

After this brief discussion on gauge issues for linearized matching theory,
let us describe the actual perturbative matching conditions (for details
see \cite{BattyeCarter01, Mukohyama00,
Mars05}). The matching theory
involves the equality of the first and second fundamental forms 
of the boundaries. To compare them they are pulled-back into a single
boundary via the identification. In perturbative matching theory
everything occurs on the abstract hypersurface $\sup_0$ diffeomorphic to
$\sup^+_0$ and $\sup^-_0$ of the background spacetime.
On $\sup^{\pm}_0$ we attach two vector
fields $Z^{\pm} = Q^{\pm} n^{(0)}_{\pm} + T^{\pm}$ whose geometric meaning has been 
discussed above. They are not know  a priori, firstly, because of the 
hypersurface gauge freedom, and secondly, because the identification 
of the boundaries $\sup^+_{\varepsilon}$ and $\sup^-_{\varepsilon}$ 
is not known a priori. To the unknowns $Z^{\pm}$ we add
four symmetric tensors $\ffffpert_{\pm}$ and $\sfffpert_{\pm}$ intrinsic to
$\sup_0$ which arise as the $\varepsilon-$derivative 
at $\varepsilon=0$ of the first fundamental form $q^{\pm}_{\varepsilon}$ 
and second fundamental form $K^{\pm}_{\varepsilon}$ of $\sup^{\pm}_{\varepsilon}$.
These tensors are intrinsic to $\sup^{\pm}_{\varepsilon}$, so before taking
$\varepsilon$-derivatives they must be pulled back onto $\sup_0$
via the hypersurface gauges (on each side). Thus, $\ffffpert_{\pm}$
and $\sfffpert_{\pm}$ are hypersurface gauge-dependent by construction. On the other
hand, their construction is fully independent of the spacetime gauge
freedom.

In order to write down their explicit expression, let 
$\gfpert{}^{\pm}$ be
the perturbed metric (i.e. the fundamental unknown in metric perturbation
theory) on each side of the background spacetime. Let also $\embed^{\pm}_0 : \sup_0 \rightarrow \mm2^{\pm}_0$ be the
embedding of the matching hypersurface on each 
region of the background spacetime. 
Let $y^i$ $(i,j,\ldots=1,\ldots,n-1)$
be a local coordinate system on $\supo$ and
define tangent vectors
$ e^{\pm}_i = \embed^{\pm}_0 {}_{\star} (\partial_{y^i})$. 
There are also  unique (up to 
orientation) unit 
one-forms $\bm{\nback_{\pm}}$ normal  to the boundaries. 
We choose them so that the corresponding vector  $\nback_{+}$ points
towards $\mm2_0^{+}$ and $\nback_{-}$ points outside of $\mm2_0^{-}$ or viceversa. 
The first and second fundamental forms of the background
are simply $\fffback{}^{\pm}\defi \embed^{\pm}_0{}^{\star} ( \gback{}^{\pm}), \quad
\sffback{}^{\pm}\defi \embed^{\pm}_0{}^{\star} (\nabla^{\pm} \bm{\nback_{\pm}} )$, where
$\nabla^{\pm}$ is the covariant derivative in 
$(\mm2^{\pm}_0,\gback{}^{\pm})$.
Given that the background configuration is already composed of
the matching of $\mmm^{+}_0$ and $\mmm^{-}_0$ through $\sup_0^+\defi\sup_0$, we already have
$\fffback{}^+=\fffback{}^-$ and $\sffback{}^+=\sffback{}^-$.
Then $\ffffpert{}^{\pm}$ and
$\sfffpert{}^{\pm}$ are defined as follows
\cite{Mukohyama00} 
\begin{eqnarray}
\ffffpert_{ij}{}^\pm&=&
{\cal L}_{T^\pm} \fffback_{ij}{}^{\pm}+
2 Q^\pm \sffback_{ij}{}^{\pm}+
e^{\pm\alpha}_i e^{\pm\beta}_j \gfpert_{\alpha\beta}{}^\pm,
\label{eq:pertfirstS}\\
\sfffpert_{ij}{}^\pm&=&
{\cal L}_{T^\pm} \sffback_{ij}{}^{\pm}-\sigma D_iD_j Q^\pm\nonumber\\
&+&Q^\pm(-\nback_\pm{}^\mu \nback_\pm{}^\nu
R^{(0)\pm}_{\alpha\mu\beta\nu}e^{\pm\alpha}_i e^{\pm\beta}_j+
\sffback_{il}{}^{\pm}{\sffback}^{l}_j{}^{\pm})
\nonumber\\
&+&\frac{\sigma}{2}\gfpert_{\alpha\beta}{}^\pm
\nback_\pm{}^\alpha \nback_\pm{}^\beta \sffback_{ij}{}^{\pm}-
\nback_\pm{}_\mu S^{(1)\pm\mu}_{\alpha\beta} e^{\pm\alpha}_i e^{\pm\beta}_j,
\label{eq:pertsecondS}
\end{eqnarray}
where $\sigma \defi \gback{}^{\pm} (\nback_{\pm}, \nback_{\pm})$,
$D$ is the covariant derivative of $(\supo,\fffback{}^{\pm})$,
$R^{(0)\pm}_{\alpha\mu\beta\nu}$ is the Riemann tensor
of $(\mm2^{\pm}_0,\gback{}^{\pm})$ and $
{S^{(1)}}_{\beta\gamma}^{\pm\alpha}\defi
\frac{1}{2} (
\nabla^{\pm}_\beta\, \gfpert{}^{\pm\alpha}_\gamma+$
$\nabla^{\pm}_\gamma\, \gfpert{}^{\pm\alpha}_\beta-
\nabla^{\pm \, \alpha}\, \gfpert{}^\pm_{\beta\gamma})$.
The first order matching conditions (in the absence of shells) 
require the  equalities
\begin{equation}
\ffffpert{}^+ = \ffffpert{}^- , \hspace{2cm}
\sfffpert{}^+ = \sfffpert{}^-.
\label{eq:fpertmc}
\end{equation}
We emphasize that $Q^{\pm}$ and $T^{\pm}$ are a priori
unknown quantities and fulfilling the matching conditions
requires {\it showing} that two vectors $Z^\pm$ exist such that
(\ref{eq:fpertmc}) are satisfied. The spacetime gauge freedom can be exploited to 
fix either or both vectors $Z^\pm$ a priori, but this should be avoided (or
at least carefully analyzed) if additional spacetime gauge choices are made, in order not
to restrict a priori the possible matchings. Regarding the hypersurface gauge, this can be used to
fix one of the vectors $T^{+}$ or $T^{-}$, but not both.
Note also that the linearized matching conditions are, by construction, spacetime
gauge invariant because, as discussed above,
the tensors $\ffffpert{}^{\pm}$, 
$\sfffpert{}^{\pm}$ are necessarily 
spacetime gauge invariant. 
In fact, it is straightforward to check explicitly that 
the right-hands sides of (\ref{eq:pertfirstS}) and
(\ref{eq:pertsecondS}) are spacetime gauge invariant (the individual terms
are not, and it is precisely the spacetime gauge dependence in $Z^{\pm}$
which makes these objects spacetime gauge invariant).
Moreover, the set of conditions (\ref{eq:fpertmc}) 
are hypersurface
gauge invariant, provided the background is properly matched,
since, as shown in \cite{Mukohyama00}, under such a hypersurface gauge transformation given by 
the vector $\zeta$ in $\supo$, $\ffffpert$ transforms as
$\ffffpert+\lie_{\zeta \, }\fffback$, and similarly for
$\sfffpert$. 

\section{Spherical symmetry: Hodge decomposition}
\label{sec:hodge}

After the previous summary on linearized matching, in this
section we introduce the second main 
ingredient for the linearized Einstein-Straus model reviewed in 
the following sections: the Hodge
decomposition on the sphere \cite{ESpert2}.  

In order to exploit the
underlying spherical symmetry of the background configuration it is
common practice to decompose the perturbations, and their related
objects and equations, in terms of scalar, vector and tensor harmonics
on the sphere. That was the procedure used 
in the seminal work on perturbations around spherical
matched background configurations, due to Gerlach and Sengupta (GS)
in \cite{GS_junction_even_79} and \cite{GS_junction_odd_79},
revisited and improved by Mart\'{\i}n-Garc\'{\i}a and Gundlach 
in \cite{Garcia_Gundlach_matching_01}.

The aim in \cite{ESpert2} was to use an alternative method, based on
the Hodge decomposition on the sphere in terms of scalars, in order to
avoid the need to deal with infinite series of objects.
In particular, the whole set of matching conditions for the
linearized Einstein-Straus model was presented in \cite{ESpert2} as a finite number
of equations involving scalars that depend on the three coordinates in
the matching hypersurface $\supo$, in contrast with an infinite number
of equations for an infinite set of functions of one variable.
It is clear that one can always go from the Hodge scalars to the
spherical harmonics decomposition in a straightforward way. However,
it is not always easy to rewrite the infinite number of expressions
appearing in a spectral decomposition in terms of Hodge scalars.

Consider
the round unit metric  $\Omega_{AB}dx^A dx^B=d \vartheta^2 + \sin^2
\vartheta d \varphi^2$, with $\eta_{AB}$ and $D_A$ denoting the 
corresponding volume form
and covariant derivative respectively, and
$(\star dG)_A= \eta^{C}{}_{A}D_C G $ the Hodge dual with respect to $\Omega_{AB}$.
Let us recall that the usual Hodge decomposition on $(\mathbb{S}^2, 
\Omega_{AB})$ states that
any one-form $V_A$ can be canonically decomposed as
$
V_A = D_A F + (\star dG)_A,
$
where $F$ and $G$ are functions on $\mathbb{S}^2$, and any
symmetric tensor $T_{AB}$ as
$
T_{AB}=D_A U_B+D_B U_A+H \Omega_{AB},
$
for some $U_A$ on $\mathbb{S}^2$, which can be in turn decomposed in terms
of scalars.

Based on this, it is convenient to define the following two functionals.
Given three scalars $\X_{\tra},\X_1$ and $\X_2$ on $(\mathbb{S}^2,\Omega_{AB})$
we define the functional one form $V_A(\X_1,\X_2)$ as
$$V_A(\X_1,\X_2) = D_A \X_1 + (\star d\X_2)_A,$$
and the functional symmetric tensor $T_{AB}(\X_\tra,\X_1,\X_2)$
as $$T _{AB}(\X_\tra,\X_1,\X_2) =  D_A V_B (\X_1,\X_2)+  D_B V_A (\X_1,\X_2)+ \X_\tra \Omega_{AB}.$$
Let us recall that the decomposition defines these $\X$'s on
$\mathbb{S}^2$ up to the kernels of the operators $V_A$ and $T_{AB}$.
We allowed for the appearance of all these kernels in \cite{ESpert2},
where their relevance (or their lack of) was already discussed.
In order to avoid 
spurious information and present a more concise review
--and also to ease the translation and
comparison with the quantities used in the previous literature
in terms of the harmonic decompositions,-- we
use the reformulation already presented in \cite{contrib_ere2012}
where the Hodge decomposition is, in fact, unique.

Indeed, in order to fix the Hodge decomposition uniquely
we define a \emph{canonical dual decomposition}
by demanding that the functions 
$\X_1, \X_2$ in $V_A(\X_1,\X_2)$ are always
orthogonal (in the $L^2$ sense on $\mathbb{S}^2$)
to 1, and the functions
$\X_1, \X_2$ in $T_{AB}(\X_\tra,\X_1,\X_2)$ 
are orthogonal
to 1 and to the $l=1$ spherical harmonics.
Schematically, we may use
  $$W_A \hodge \X_1, \X_2 \qquad \mbox{ to indicate } \qquad W_A=V_A(\X_1, \X_2)$$
and $$W_{AB} \hodge \X_{tr},\X_1, \X_2 \qquad \mbox{ when }\qquad
W_{AB}=T_{AB}(\X_{tr},\X_1,\X_2).$$

We will use the following notation: given any function
$f$ on $\mathbb{S}^2$ we define
\[
f_{||0}\defi Y_0\int_{\mathbb{S}^2}f\, Y_0 d \Omega^2,\qquad
f_{||1}\defi Y_1\int_{\mathbb{S}^2}f\, Y_1 d \Omega^2\left(=\sum_mY^m_1\int_{\mathbb{S}^2}f\, Y^m_1 d \Omega^2\right),\]
so that $f-f_{||0}$ is orthogonal to the $l=0$ harmonics 
$Y_0$ and $f-f_{||1}$ is orthogonal to the $l=1$ harmonics $Y_1^m$.

Note finally that the Hodge decomposition in terms of scalars involves two types
of objects depending on their behaviour under reflection on the sphere.
The scalars with subscripts $1$ and $\tra$ remain unchanged under reflection,
and are typically called longitudinal,
even or polar quantities, while those with subscripts $2$ change, and
correspond to the transversal, odd or axial quantities.

Let us consider now the general spherically symmetric spacetime
$\mmm=M^2\times \mathbb{S}^2$ with metric
$
g_{\alpha\beta}=\omega_{IJ} \oplus r^2 \Omega_{AB},
$
so that $(M^2,\omega_{IJ})$ is a 2-dim Lorentzian space and
$r>0$ a function on $M^2$.
The dual in $(M^2, \omega_{IJ})$ will be indicated by $*$
and the covariant derivative by $\nabla$.

We can now proceed to decompose any
one-form (vector) or symmetric two-tensor on $\mmm$ by first taking
the part orthogonal to the sphere and then apply the Hodge
canonical decomposition to the part tangent to the sphere.
In particular, given a normalized timelike one-form $u_\alpha$ orthogonal to the
spheres, its corresponding one-form $u_I$ on $(M^2,\omega_{IJ})$
(defined by $u_\alpha=(u_I,0)$) 
can be used to construct a convenient orthonormal basis
$\{u_I,m_I\}$ so that $\omega_{IJ}=-u_Iu_J+m_I m_J$
(this is $u_Iu^I=-1$ and $m_I\defi *u_I$), and consider then
the one-form on $\mmm$ defined by $m_\alpha\defi(m_I,0)$.
Given any vector $V_I$ 
we will simply denote by $V_u$ and $V_m$ the contractions
$u^I V_I$ and $m^I V_I$ respectively.

We apply now this decomposition to encode the objects that will describe the (first order) perturbation
of a background consisting of
two spherically symmetric regions $(\mm2^+,\gback{}^+)$ and $(\mm2^-,\gback{}^-)$
matched across corresponding spherically symmetric boundaries $\supo^+$ and  $\supo^-$.
At each side $\pm$ (we avoid the use of $\pm$ just now for clarity)
the metric perturbation tensor
$\gfpert_{\alpha\beta}$ gets thus decomposed as
\begin{eqnarray}
  &&\gfpert_{IJ} = \Z_{IJ},\nonumber\\
  &&\gfpert_{I A} \hodge \Z_{I1}, \Z_{I2},\nonumber\\
  &&\gfpert_{AB} \hodge  \ZStwo_{\tra}, \ZStwo_1, \ZStwo_2,\label{eq:g_a_Z}
\end{eqnarray}
where $\Z_{I1}$ and  $\Z_{I2}$ are two one-forms defined on $M^2$,
and analogously for any symmetric tensor.
The deformation vector $Z_\alpha$, defined on $M$ at points
on $\sup$, is decomposed as $$Z_\alpha\to
     \  \{Z_I \to Q  , T\}\oplus
        \{Z_A \hodge \mathcal{T}_1,\mathcal{T}_2\},
     $$
whereby the part $Z_I$ gets decomposed, in turn,
onto the normal and tangential parts to $\supo$, $Q$ and $T$ respectively.
Given 
the (first order)
perturbation tensor and the deformation vector at either $\pm$ side of the matching hypersurface,
one can calculate the symmetric tensors $\ffffpert_{ij}$ and $\sfffpert_{ij}$, i.e. the ``perturbed first and second fundamental forms'', using (\ref{eq:pertfirstS})-(\ref{eq:pertsecondS}).
Recalling now that $\ffffpert_{ij}$ and $\sfffpert_{ij}$ are defined on $(\supo,\fffback_{ij})$
and that $\supo$ at either side are tangent to the spheres $\{\theta,\phi\}$,
let us denote by $\lambda$ the parameter that follows the
direction on $\supo$ orthogonal to the spheres in order to decompose
$\ffffpert_{ij}$ and $\sfffpert_{ij}$ into $\ffffpert_{\lambda\lambda}$, $\sfffpert_{\lambda\lambda}$, plus 
\begin{equation}
  \ffffpert_{\lambda A} \hodge F^q, G^q,\qquad
  \ffffpert_{AB} \hodge  H^q, P^q, R^q,
\end{equation}
and
\begin{equation}
  \sfffpert_{\lambda A} \hodge F^k, G^k,\qquad
  \sfffpert_{AB} \hodge  H^k, P^k, R^k.
\end{equation}
Note that all these functions are scalars on the sphere that depend only on $\lambda$.
The first order matching conditions (\ref{eq:fpertmc}) are therefore equivalent to
\begin{eqnarray}
  &&\ffffpert_{\lambda\lambda}{}^+=\ffffpert_{\lambda\lambda}{}^-,
  \quad F^q_+=F^q_-, \quad G^q_+=G^q_-, \quad H^q_+=H^q_-,\quad P^q_+=P^q_-, \quad R^q_+=R^q_-, \nonumber\\
  &&\sfffpert_{\lambda\lambda}{}^+=\sfffpert_{\lambda\lambda}{}^-,
  \quad F^k_+=F^k_-, \quad G^k_+=G^k_-, \quad H^k_+=H^k_-,\quad P^k_+=P^k_-, \quad R^k_+=R^k_- .
\label{eq:matching_scalars}
\end{eqnarray}
Except for the simplification of the kernels by the canonical decomposition,
these are the  
linearized matching conditions presented in \cite{ESpert2}.

\subsection{Gerlach and Sengupta (GS) 2+2 formalism}
Let us emphasize again that the conditions (\ref{eq:fpertmc}), and thus (\ref{eq:matching_scalars}),
concern quantities defined on $\supo$ and are therefore independent on the coordinates
used in $\mm2^+$ and $\mm2^-$ for their computation. 
These quantities are thus, on the one hand, spacetime gauge independent by construction.
Moreover, as discussed in Section \ref{sec:overview}, the \emph{equations} are also
hypersurface gauge independent.
All this makes it unnecessary the use of gauge invariant quantities in order
to establish the perturbed matching conditions. Having said that, however, 
the use of gauge independent quantities turns out to be convenient in the end, mostly
when one eventually wants to impose the Einstein field equations.
As shown in the works
\cite{GS_junction_even_79}, \cite{GS_junction_odd_79},
\cite{Garcia_Gundlach_matching_01},
the use of spacetime gauge invariants
is very convenient in order to combine the
Einstein field equations with the perturbed matching conditions.

One can proceed by constructing gauge invariants in terms
of Hodge scalars using analogous expressions to those in
the harmonic decomposition constructed in
\cite{GS_junction_odd_79,GS_junction_even_79,Garcia_Gundlach_matching_01}.

Let us now concentrate on the odd (axial) sector.
The odd gauge invariant quantities are encoded in the vector \cite{ESpert3} (cf. \cite{GS_junction_odd_79})
     \[
     \Kas_{I}\defi \Z_{I2} -\nabla_I \ZStwo_2 + 2 \ZStwo_2 r^{-1}\nabla_I r.
     \]
Note that  $\Kas_I$, as defined above, contains $l=1$ harmonics,
from $\Z_{I2}$, but only the $l\geq 2$ sector is gauge invariant.
In other words, the part of $\Kas_I$ orthogonal to $Y^1$ (i.e.
$\Kas_{I}  - \Kas_{I}{}_{||1}$
in the notation above)
is the gauge invariant part.
Once the orthonormal basis $\{ u_I, m_I\}$ has
been identified at both sides, the odd sector of the linearized matching,
which corresponds to the set of equations
\begin{equation}
  G^q_+=G^q_-,\qquad
R^q_+=R^q_-,\qquad G^k_+=G^k_-,\qquad R^k_+=R^k_-,
\label{eq:odd_matching}
\end{equation}
  in
(\ref{eq:matching_scalars}), is equivalent in the $l\geq 2$ sector (the part orthogonal to $l=0$ and $l=1$) to
\cite{ESpert3} (cf. \cite{GS_junction_odd_79})
\begin{equation} 
\Kas^+_u\eqq \Kas^-_u,\qquad \Kas^+_m\eqq \Kas^-_m,\qquad
     * d(r^{-2}\bm\Kas^+)\eqq * d(r^{-2}\bm\Kas^-)
\label{eq:firstMC_Kas}
\end{equation}
plus an equation for $\mathcal{T}_2^+-\mathcal{T}_2^-$.

\section{Linearized Einstein-Straus model: matching conditions}
\label{sec:linearizedmatching}
We are ready to consider the linearized matching
of the perturbed Schwarzschild and FLRW spacetimes, as it was
analyzed in \cite{ESpert2}.
Take the Einstein-Straus model as described in Section \ref{sec:ESmodel}:
the FLRW geometry in cosmic time coordinates (\ref{FLmetric}),
the Schwarzschild in standard coordinates  (\ref{SCHmetric}),
and the background matching hypersurface $\supo$ described by 
$\supo^\FL: \{ t=t, R = R_0 \}$ and
$\supo^\stc: \{ T=T_0(t), r = r_0(t) \}$ respectively, where
$T_0(t)$ and $r_0(t)$ satisfy (\ref{matchingODE})
and the angular part is again ignored.

The orthonormal basis we take on the Lorentzian space
orthogonal to the spheres is formed
by $\bm u= dt$, $\bm m=a dR$. Note that this choice corresponds to the tangent and normal
forms to $\supo$ at $\supo$, respectively. More precisely, $\bm m$ is chosen to be $\bm\nback$
at points on $\supo$. On the Schwarzschild side we have $u|_{\supo}=\dot T_0\partial_T+\dot r_0\partial_r$
 and $\bm m|_{\supo} \defi \bm\nback=-\dot r_0 dT+\dot T_0dr$.

Take the first order perturbations of FLRW,
in no specific gauge,
formally decomposed into the usual scalar, vector and tensor (SVT)
modes, i.e.\footnote{Note that in \cite{ESpert2}, the FLRW geometry was written in conformal time, while proper time is used here.} (Latin indices $a,b,c$ are used for tensors on $(\M,g_{\M})$)
\begin{eqnarray*}
  &&\gfpert_{tt}{}^+=-2\Psi\qquad
  \gfpert_{ta}{}^+=aW_a\\
  &&\gfpert_{ab}{}^+=a^2(-2\Phi \gamma_{ab}+\chi_{ab})
\end{eqnarray*}
with $$W_a=\partial_a \wzero+\tilde W_a,\qquad
      \chi_{ab}=
        (\nabla_{a}\nabla_{b}-\frac{1}{3}\gamma_{ab}\nabla^2)\cchi
        +2\nabla_{(a}\cy_{b)}+\ppi_{ab}$$
satisfying the constraints
 $$\nabla^a\cy_a=\ppi_{a}^{~a}=0,\qquad \nabla^a\ppi_{ab}=0,\qquad \nabla^a\tilde W_a=0.$$
The canonical Hodge decomposition is then used to encode the part
tangent to the spheres into $\mathbb{S}^2$ scalars in the following schematic way \cite{ESpert2}:

\noindent
\emph{vector} $$\tilde W_a\to\tilde W_R\oplus\{\tilde W_A  \hodge \wone,\wtwo\},
\qquad \cy_a\to\cy_R\oplus\{\cy_A  \hodge \yone,\ytwo\},$$

\noindent
\emph{tensor} $$\ppi_{RA}\hodge\qone,\qtwo,\qquad
      \ppi_{AB}\hodge \ch,\uone,\utwo.$$
All in all, encoding $\gfpert{}^+$ using the SVT modes together with the
Hodge decomposition leaves us with 15 SVT-Hodge
quantities,
corresponding to the
scalar modes $\{\Psi,\Phi,\wzero,\cchi\}$,
vector modes $\{\tilde W_R,\wone,\wtwo,\cy_R,\yone,\ytwo\}$
and tensor modes $\{\qone,\qtwo,\ch,\uone,\utwo\}$,
all scalars
on $\mathbb{S}^2$, not all independent
due to the previous constraints, and, on the other hand, \emph{not unique}.
Consider, in particular, the only 4 scalars we have in the
odd sector;
vector modes $\{\wtwo,\ytwo\}$ and
tensor modes $\{\qtwo,\utwo\}$. As discussed in the previous
section, only two gauge invariant quantities exist in the
odd sector. These correspond to the two components
of the gauge invariant odd vector (and for $l\geq2$), $\Kas^+_I$, which given the above construction in terms of the
SVT-Hodge quantities, read
\cite{contrib_ere2012,ESpert3}
$$
\Kas^+_u=a\left(\wtwo-a(\dot\utwo+\dot \ytwo)\right),\qquad
\Kas^+_m=a\left(\qtwo-\derxiutwo
            +2\utwo\dersig\sig^{-1}\right).$$

Consider now the stationary and axially symmetric vacuum perturbations
in the Weyl gauge.
They can be described in terms of two functions
$U^{(1)}(r,\theta)$ and $A^{(1)}(r,\theta)$, which correspond, basically,
to the perturbation of the gravitational Newtonian potential
and the rotational perturbation, respectively.
The perturbation tensor reads \cite{ESpert2}
\begin{eqnarray}
      \gfpert{}^{\Sch}&=& 
      - 2 \left (1- \frac{2 m}{r} \right ) 
      \left ( U^{(1)} dt^2+   {A^{(1)}} dt d \phi \right )\nonumber\\
      &&- 2 r^2 \sin^2 \theta  {U^{(1)}} d\phi^2 
      + 2 \left (k^{(1)} - U^{(1)} \right ) \left[ \left(1 - \frac{2 m}{r}\right)^{-1}dr^2  + r^2 d \theta^2 \right].\label{eq:pert_Sch}
    \end{eqnarray}
Note that, when using the full set of the Einstein field equations for
vacuum, the function $k^{(1)}$ is determined up to quadratures once
$U^{(1)}(r,\theta)$ and $A^{(1)}(r,\theta)$ are found.
We stress,
however, that the vacuum equations, although indicated,
were not imposed in the perturbed
Schwarzschild region (nor in the perturbed FLRW region) in
\cite{ESpert2}, and therefore the results found there are purely
geometric.

Instead of working with $A^{(1)}(r,\theta)$,
it is convenient to use an auxiliary function, $\cg$,
defined by $A^{(1)} \defi  \sin \theta \partial_\theta\cgsup$.
In terms of the Hodge decomposition of the perturbation tensor
(\ref{eq:g_a_Z}), applied to (\ref{eq:pert_Sch}), we have
$\cgsup= -\left (1- \frac{2 M}{r} \right )^{-1}\Z^-_{T2}$.
Another auxiliary function $\mathcal{P}$ can be introduced for $k^{(1)}$.

The whole set of matching conditions (\ref{eq:matching_scalars})
for the linearized Einstein-Straus model, both in the odd and
even sectors,
were found in \cite{ESpert2} in terms of these functions
$U^{(1)}(r,\theta),\cg$ and $\mathcal{P}$ in the Schwarzschild side
together with the above 15 SVT-Hodge scalars
describing the FLRW perturbation.
The whole set is too long to be included here,
but in order to review the main results in \cite{ESpert2}
only the odd sector of the linearized matching is needed.

The odd sector of the linearized matching (\ref{eq:odd_matching})
projected to the part
orthogonal to the $l=0$ and $l=1$ harmonics 
can be rewritten as the following three relations \cite{ESpert2}
\begin{eqnarray}  
\label{eq:W2}
&&{{\cal W}_2} -a [{\dot{\cal U}_2} + {\dot {\cal Y}_2}
]\eqq   -   \cgsup \dersigo \asig^{-1},\\
\label{eq:Q2}
&&\qtwo -
\derxiutwo 
+2\sigo^{-1}\dersigo\, 
\utwo 
\eqq -\cgsup \asig^{-1}\asigdot \sigo,\\
\label{eq:dW2}
&&{{\cal W}_2}'-a{\frac{d}{d t}}[{({\cal U}_2}' + {{\cal Y}_2}')|_{\supo} 
]
\eqq 
 \cgsup \frac{\sigo^3 \asig\curv -3M}{\sigo^2 \asig^2}
+\dercgsupr(\sigo^2\curv-1),
\end{eqnarray}
plus an equation for the difference $\mathcal{T}_2^\FL-\mathcal{T}_2^\Sch$.
The three equations above can be shown \cite{ESpert3} to
correspond indeed to (\ref{eq:firstMC_Kas}),
taking into account that 
$\bm\Kas^\Sch=-\left (1- \frac{2 M}{r} \right )\cgsup dT$.
These equations were presented in \cite{ESpert2} in full,
including the parts lying on the $l=0$ and $l=1$ harmonics.
There, a series of kernels inherent to the usual
Hodge decomposition where the responsible for the usual freedom
found in the $l=0$ and $l=1$ sectors when using harmonic
decompositions.
By using the \emph{canonical} Hodge decomposition
introduced in Section \ref{sec:hodge}, we can have a better
control of that freedom, and understand its nature,
getting rid of the spurious terms.
Indeed, by doing that it can be shown \cite{ESpert3} that
the projection of the linearized matching on the $l=1$ harmonics
--on the $l=0$ it is trivial, since all scalars in the odd sector
are orthogonal to $l=0$-- gives 
\begin{align}
&\dot\asig\asig^2\sigo^2\left[\mathcal{W}'_2{}-\asig\frac{d}{d t}\mathcal{Y}'_2|_{\supo}
 -2\sigo^{-1}\dersigo \left(\mathcal{W}_2{}-\asig\frac{d}{d t}\mathcal{Y}_2{}\right)\right]_{||1}\nonumber \\
&\qquad\eqq\asig(2M-\asig\sigo)\dot\cgsup_{||1}+2\dot\asig(\asig\sigo-3M)\cg_{||1}
\label{eq:cg_l1}
\end{align}
while $(\mathcal{T}_2^\FL-\mathcal{T}_2^\Sch)_{||1}$ is free.

The first consequence of the above equations is that if the FLRW
remains unperturbed then the stationary region must be static in the
range of variation of $r_0(t)$:
equations (\ref{eq:W2})-(\ref{eq:dW2}) imply that the part of
$\cgsup$ orthogonal to $l=1$ vanishes, and (\ref{eq:cg_l1}) implies
that
$
\cgsup_{||1}=C a^3/(2M-\asig\sigo),
$
where $C$ is a constant. Therefore $\cgsup=C a^3/(2M-\asig\sigo)\cos\theta$,
and thus\footnote{This equation was obtained in \cite{ESpert2}
by deriving the necessary constraints for the aforementioned kernels.}
$$A^{(1)}|_{\supo} = C  \sin^2 \theta a^3/(2M-\asig\sigo).$$
As shown in \cite{ESpert2}, this implies that the perturbed
spacetime is static in the
range of variation of $r_0(t)$.

This result
generalizes that in \cite{Nolan_Vera_05} because now the matching
hypersurface does not necessarily keep the axial symmetry.  Therefore,
the only way of having a stationary and axisymmetric vacuum
arbitrarily shaped (at the linear level) region in FLRW is to have the
Einstein-Straus model.
Let us remark again that by the interior/exterior duality,
this result also implies that a piece of FLRW, irrespective of its shape
and its relative rotation with the exterior,
cannot describe the interior of Kerr.

\subsection{Constraint on FLRW}
Another interesting consequence of the matching conditions is that the combination of (\ref{eq:W2})
and (\ref{eq:Q2}) produces one equation that involves only quantities
in FLRW \cite{ESpert2}
\begin{eqnarray*}
   \frac{\asigdot\sigo}{\asig\dersigo}\left(\wwtwo\right) \eqq \qtwo - \derxiutwo 
  +2\sigo^{-1}\dersigo \utwo, 
\end{eqnarray*}
and thus constitutes a \emph{constraint} in FLRW.
Recalling that this equation is meant to be orthogonal to $l=1$
(and vanishes identically if projected on $l=0$),
this constraint implies that if the perturbed FLRW contains
vector modes  with $l\geq 2$ harmonics on $\supo$, then it must contain
also tensor modes there. Since, as we have just seen above,
the existence of a rotation in the stationary region 
implies the existence of, at least, vector perturbations in FLRW,
then both vector and tensor modes must exist on $\supo$.

It must be stressed that, as demonstrated in \cite{Bicak_Katz_Lynd_07} (and
references therein), there exist configurations of FLRW
linear perturbations containing only vector perturbations
which vanish identically inside a spherical surface. Such
configurations are compatible with the results reviewed
here, since that interior region is FLRW and the above
constraints do not apply. A completely different matter is
the embedding of a Schwarzschild spherical cavity (or a
vacuum perturbation thereof) into any such model: the
Schwarzschild cavity cannot reach the perturbed FLRW
region, as otherwise the constraints above would require
that tensor perturbations are also present (at least near the
boundary of the Schwarzschild cavity).

There also exists a Einstein-Straus perturbative model by Chamorro
\cite{Chamorro88}
consisting on small rotation Kerr vacuole within 
a perturbed FLRW. Again, the constraint above
does not apply to this model because there are no $l\geq 2$ modes there.

The fact that tensor modes must exist near $\supo$ once
some rotation with $l\geq 2$, whatever small, exists in the stationary region,
may indicate the existence of some kind of gravitational waves
on FLRW near $\supo$. In order to analyze further this issue
one needs to take the Einstein field equations
into consideration. That is the purpose
of our work in preparation \cite{ESpert3}, some result of which
will appear in the proceedings of the ERE2012 meeting.

\section{Conclusions and outlook}
\label{sec:conclusions}

This paper is concerned with the difficulties that the
Einstein-Straus model encounters. A fundamental one refers to its high
level of rigidity and the impossible generalization to non-spherical
symmetry if the bound system is required to be time independent 
so as to retain the property that cosmic expansion does not affect the local 
systems.  Moreover, if one views the model within the LTB class with a step
function density profile, the model is unstable to perturbations
\cite{Sato84},  \cite{LakePim85}, cf. also the discussion in 
\cite{Krasinski}. 
The rigidity result so far requires
either  stationarity and axial symmetry or staticity. An interesting
open problem would be to relax the conditions and assume 
only stationarity.

Despite these difficulties, the Einstein-Straus model has played and plays 
a very important role in cosmology in
different areas or research, most notably on the influence (or lack thereof)
of the cosmic expansion on local systems, or in the problem of averaging
in cosmology at least on an observational level
(see e.g. \cite{RelCosm}). 
The model is still
widely used as textbook explanation of the lack of influence of
the cosmic expansion on astrophysical systems and, in fact, there are not many
known alternatives (a notable exception is the 
McVittie model \cite{McVittie}, \cite{Jarnefelt1940}, \cite{Jarnefelt1942}
which is also spherically symmetric
and mimics the geometry of Schwarzschild at small
scales while approaching a FLRW model at long distances, and which has
been studied thoroughly, see \cite{Carrera} and references
therein). Concerning the use of the Einstein-Straus model
on observational cosmology, mainly by studying lensing effects,
its generalizations 
have systematically consisted in keeping spherical symmetry and allowing for
some dynamics in the interior. The prominent example
here consists of LTB regions inside a FLRW universe (see references in
\cite{RelCosm}). 

An important ingredient for the rigidity of the Einstein-Straus model
is the large symmetry of the FLRW background. It is therefore an
interesting problem to analyze how the model gets modified in the presence
of cosmic perturbations.
In a conservative approach, one still wants to keep
the main properties (stationarity of a region inside a cosmological model)
as far as possible and analyze the possible departures from the model
in more realistic situations.
Moreover, by studying perturbed Einstein-Straus models one seeks going beyond
the problem of the influence of cosmic expansion on local systems,
and tackle the problem of the influence of general cosmic dynamics.
Surprisingly, it turns out that the existence of static (stationary) regions
does impose conditions on the cosmic perturbations, at least
near the boundary. Whether this is a real effect or simply an indication that
the interior region should not be kept stationary remains to be seen.

Another implication of the perturbed Einstein-Straus model is that
extra care is required in the standard decomposition 
of metric perturbations in terms of 
scalar, vector and tensor modes. As discussed above,
any rotation in the vacuole implies
necessarily the  presence of both vector and tensor modes in the cosmic
perturbations. It is interesting to analyze  whether these tensor modes
could represent cosmic gravitational waves.
Some preliminary results along
these lines have already been presented in \cite{contrib_ere2012}. A detailed
and more complete approach will appear elsewhere.

Another interesting future line of research is to allow for non-stationary
perturbations in the vacuole and study the transmission of gravitational
waves from the cosmic region to the bound system.

\section{Acknowledgements}

MM acknowledges financial support under the projects FIS2012-30926 (MICINN)
and P09-FQM-4496 (J. Andaluc\'{\i}a---FEDER). 

FM thanks the warm hospitality from Instituto de F\'isica, UERJ, Rio de Janeiro, Brasil, projects PTDC/MAT/108921/2008 and CERN/FP/123609/2011 from Funda\c{c}\~ao para a Ci\^encia e a Tecnologia (FCT), as well as CMAT, Univ. Minho, for support through FEDER funds Programa Operacional Factores de Competitividade (COMPETE) and Portuguese Funds from FCT within the project PEst-C/MAT/UI0013/2011.

RV thanks the kind hospitality from the Universidad de Salamanca,
where parts of this work have been produced,
and financial support from project IT592-13 of the Basque
Government, and FIS2010-15492 from the MICINN.


\end{document}